\documentclass[prl,twocolumn,
longbibliography,
showpacs,
aps
]{revtex4-1}
\usepackage[english]{babel}      
\usepackage{graphicx}
\usepackage{amsmath}
\usepackage{amsfonts}
\usepackage{amssymb}
\usepackage[colorlinks=true,linkcolor=blue,urlcolor=blue,citecolor=blue]{hyperref}
\usepackage{color}
\usepackage{lipsum} 
\usepackage{float}
\usepackage{subfigure}
\usepackage{dcolumn}
\usepackage{dsfont}
\usepackage{placeins} 
\usepackage{bm}
\newcommand{\im}{\mathrm{i}}
\newcommand{\e}{\mathrm{e}}
\usepackage{nicefrac}
\usepackage[normalem]{ulem}

\begin{document}


\title{Tunneling resonances through periodically driven quantum dots}

\author{Jan~Mathis~Giesen}
	\affiliation{Fachbereich Physik and Landesforschungszentrum OPTIMAS, University of Kaiserslautern-Landau, 67663 Kaiserslautern, Germany}
	
\author{Daniel~Weber}
	\affiliation{Fachbereich Physik and Landesforschungszentrum OPTIMAS, University of Kaiserslautern-Landau, 67663 Kaiserslautern, Germany}

 \author{Sebastian~Eggert}
	\affiliation{Fachbereich Physik and Landesforschungszentrum OPTIMAS, University of  Kaiserslautern-Landau, 67663 Kaiserslautern, Germany}


\date{\today}

\begin{abstract}
 Periodic driving of quantum dots is analyzed as a basis for developing dynamic switching devices.  We study transport through 
periodically modulated energy levels which are coupled  to  leads via tunneling coefficients.  Utilizing Floquet theory a full analytic solution is found in terms of continued fractions, enabling us to efficiently calculate and analyze the transmission through the quantum dot in relevant parameter regimes.  
By considering levels at higher energy outside the spectrum of the transmitted particles a resonant switching effects is identified, where a  very small oscillating control signal on
a weakly connected quantum dot can induce
perfect transmission.  We also find closed form expressions using Bessel functions in the limit of small tunnel couplings. 
The results predict and explain resonant tunneling in nano-electronic devices
as well as in corresponding setups using magnonic
systems, photonic waveguides, or ultra-cold gases in optical lattices.
\end{abstract}

\maketitle
{\it Introduction.} The demand for more efficient switching devices has inspired physics research to develop ever smaller "quantum dots" down to molecular electronics and molecular junctions \cite{Aviram1974,Reed1997,Cui2001,PhysRevB.63.045416,PhysRevB.65.113410,Galperin2008,Shen2010, Garrigues2016,SONG2020110514,PhysRevB.109.075156,Yang2023, Nitzan2003,Metzger2015,Xiang2016,HAN2024100517}.  
 A particular promising area of 
research is {\it photon-assisted tunneling}  \cite{PLATERO20041,Yuan2014,Sun1997,Jauho1994,Qin2001,Yang2014,Shafranjuk2008,Yang2017,Drexler1995, PhysRevB.50.2019,Kouwenhoven1994,Lee2019,Kouwenhoven1991,Blick_1995,Inarrea_1997,PhysRevLett.78.1536,PhysRevB.65.113304,Shibata2012,PhysRevB.107.115165,PhysRevB.97.195423,10.1063/5.0184978}, where time-periodic fields are used to excite electrons  to and from energy levels which normally cannot participate in the transport.   A local gate is not necessarily required when using electromagnetic waves, which 
may help miniaturization.
Another advantage is that in addition to the underlying energy level $\mu_0$, there are two more adjustable parameters: amplitude $\mu$ and frequency $\omega$ of the applied field.  
It may be viewed as a disadvantage that  photon-assisted tunneling appears to require one photon for each electron to be excited and transported, which translates into sizeable
radiation.  However, as we will show here,  this is not true for a coherent quantum solution, where effective "Floquet" levels are dynamically created which assist the tunneling.
We now develop an analytic non-equilibrium solution of the general driven model and predict when it is possible to switch on {\it perfect transmission} even in the 
limit of very small oscillating fields $\mu$ and tunneling amplitudes $J'$.  

\begin{figure}[t]
    \centering
    \includegraphics[width=0.48\textwidth]{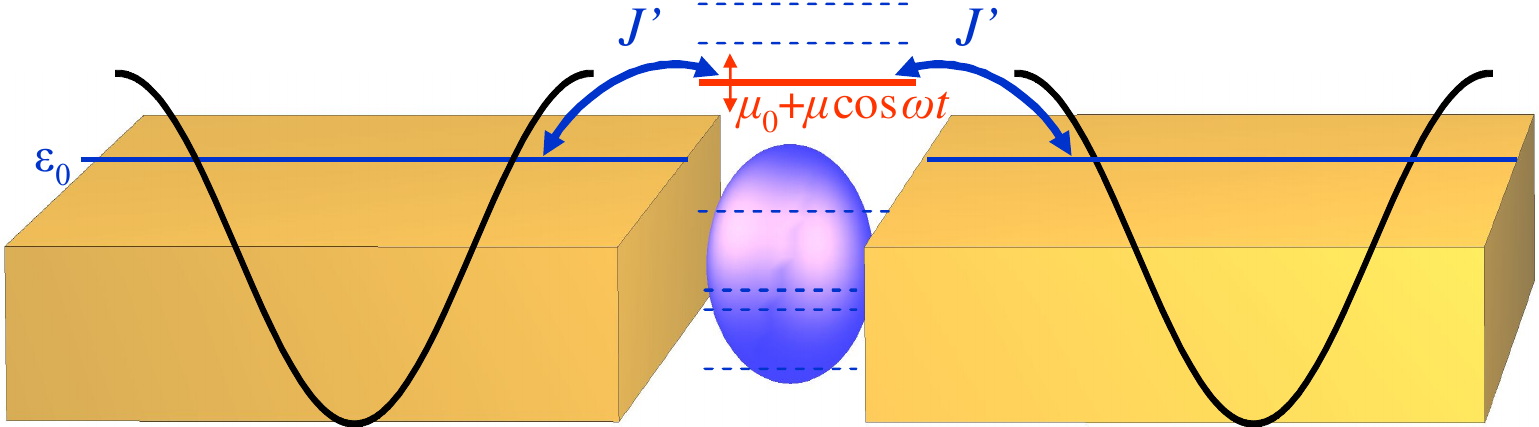}
    \caption{Schematic setup of transport through a periodically driven quantum dot.\vspace{-0.7cm} }
    \label{fig:sketch}
\end{figure}

A generic
setup is depicted in Fig.~\ref{fig:sketch} where the conduction between leads, gates, or tunneling tips is governed by the energy structure of a quantum dot.
Tunneling can occur via each of the levels, which therefore form independent "channels" that can be manipulated by oscillating fields or even vibrations.
Previous works have considered transport through oscillating barriers with good connections $J'\approx J$ \cite{PhysRevB.93.180301,Reyes_2017,Thuberg2017a,PhysRevB.96.104309,Huebner2021,PhysRevA.108.023307},
where a large {\it reduction} of  transmission  and filtering was observed for very small driving amplitudes $\mu$.  
The opposite effect of 
large {\it increase} of transmission for very weak connection $J'\ll J$ would be very valuable for switching, since a single channel could open up perfect transport in this case, which  
is generically difficult to achieve by a small driving amplitude.
We now seek to overcome this limitation by using Floquet theory  \cite{Floquet,shirley, eckardt, Holthaus} to derive 
an analytic solution for the transmission coefficient in the general setup.
Using this result it is possible to search for large transmission {maxima} in the entire parameter space.
By including oscillating energy levels of the quantum dot {\it outside}  the band it is possible to open 
the door for perfect resonant tunneling even in the limit of  very small driving amplitudes $\mu$ and connections $J'$.

{\it Model.} As shown in Fig.~\ref{fig:sketch}, the transport is assumed to be governed by incoming and outgoing particles between two leads at a given energy $\epsilon_0$,  with a negligible voltage drop.
The non-equilibrium situation in the presence of radiation allows to involve levels $\mu_0$  away from $\epsilon_0$ outside and inside the band.
In particular, it turns out that the locations of the band edges play an important role.  We use a cosine-like band-structure, which is conveniently 
implemented by a tight-binding model with hopping $J$ along the direction of transport and  a tunneling coefficient $J'$ to the quantum dot 
\begin{eqnarray}
H \!& =&\!  -J\! \!\sum_{j\neq-1,0} \!\!c_j^\dagger c_{j+1}{\phantom{\dagger}}   \!\! 
\! -\!J'\!\! \sum_{j=-1,0}\!\! (c_{-1}^\dagger c_{0}{\phantom{\dagger}}\! +\! c_{0}^\dagger c_{1}{\phantom{\dagger}}\!\!\!) + h.c. \nonumber \\
&   & +  \left(\mu_0+\mu \cos (\omega t)\right) c_0^\dagger c_{0}{\phantom{\dagger}}.
\label{model}
\end{eqnarray}
The left and right leads for $j\!<\!0$ and $j\!>\!0$, respectively, are assumed to be three-dimensional, but 
the other two directions as well as the spin degree of freedom do not influence the transmission and have been omitted in the model.    
Here, the quantum dot is represented by the central site $j\!=\!0$, i.e.~the transmission can be calculated for any given energy level $\mu_0$ on the dot separately.
  Each unoccupied level $\mu_0$  can be treated 
as an independent channel as long as the incoming energy $\epsilon_0$
is not close to the energy  $\mu_0$.   
It should be noted that interactions effects arise from double occupancies for levels 
$\mu_0 \!\sim\! \epsilon_0\pm V$ within a finite voltage range $V$ \cite{wind, dolan}, which is an
interesting situation for the static case \cite{wingreen}
as a resource of single electron transistors  \cite{RevModPhys.64.849}. 
While this remains a very active research field, we are now interested if  switching 
can be achieved from higher {\it unoccupied} quantum dot levels  using oscillating fields.  
We would like to emphasize that the general model in Eq.~(\ref{model}) is relevant for a range of other experimental systems where (quasi)particles are transported through 
an oscillating potential, such as magnonic systems \cite{PhysRev.58.1098,10.1063/1.3499236,Shen_2018}, photonic waveguides \cite{waveguide, GARANOVICH20121,PhysRevResearch.3.013260}, or ultracold gases in optical lattices \cite{Bloch2005}.

\begin{figure}[t]
    \centering
     \includegraphics[width=0.49\textwidth]{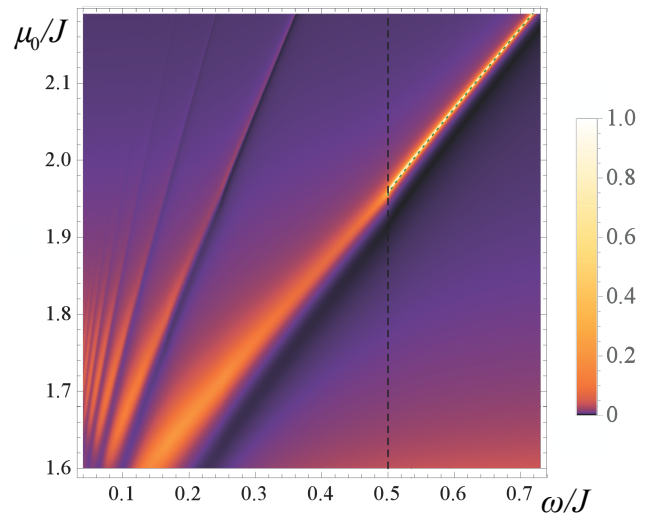}
    \caption{Transmission $|t_0^2|$  for incoming energy $\epsilon_0\!=\!1.5J$ as a function of driving frequency $\omega$ and the energy level $\mu_0$ of the quantum dot at driving amplitude $\mu\!=\!0.3J$ and coupling $J'\!=\!0.15J$.   The black dashed line is the band edge $\omega\!+\!\epsilon_0\!=\!2J$
    and the dotted green line is the prediction in Eq.~(\ref{res}). } 
    \label{fig:cplot_mu0} \vspace{-.5cm}
\end{figure}

{\it Floquet solution.} In order to determine the transmission coefficient we use a Floquet \cite{Floquet,shirley, eckardt, Holthaus} ansatz $|\psi(t)\rangle\!=\!\mathrm{e}^{-\mathrm{i}\epsilon t}|\phi(t)\rangle$ with periodic Floquet modes $|\phi\left(t\!+\!T\right)\rangle\!=\!|\phi(t)\rangle$ 
to solve the Schr\"odinger equation $(H(t)\!-\!\mathrm{i}\partial_t)|\psi(t)\rangle\!=\!0$ for the time periodic 
Hamiltonian in Eq.~(\ref{model}), where $T\!=\!\frac{2\pi}{\omega}$ is the driving period and $\epsilon$ the quasi-energy fulfilling 
\begin{align}
    (H(t)-\mathrm{i}\partial_t)|\phi(t)\rangle=&\epsilon |\phi(t)\rangle.
\end{align}
Writing the Hamiltonian and the Floquet modes as their spectral decomposition $H(t)\!=\!\sum_n \e^{-\im \omega n t}H_n$ and $|\phi(t)\rangle\!=\!\sum_{n} \mathrm{e}^{-\mathrm{i}n\omega t}|\phi_n\rangle$ we get
\begin{align}
    H_0|\phi_n\rangle + H_1(|\phi_{n+1}\rangle+|\phi_{n-1}\rangle)=(\epsilon+n\omega)|\phi_n\rangle
\label{eq:H_floquet}
\end{align}
where $H_1\!=\!\frac{\mu}{2}c_0^\dagger c_0$ and $H_0$ is the static part of Eq.~(\ref{model}). 
To calculate the transmission coefficient $t_0$ we assume an incoming wave with wave vector $k_0$
\begin{align}
\nonumber
   |\phi_n\rangle =& \sum_{j<0}(\delta_{n,0}\e^{\im k_0j}+r_n\e^{-\im k_nj})c_j^\dagger|0\rangle \\
   &+\xi_nc_0^\dagger|0\rangle + \sum_{j>0}t_n\e^{\im k_nj}c_j^\dagger|0\rangle
\label{eq:Fansatz}
\end{align}
where $-2J\cos(k_n)\!=\!\epsilon_0+n\omega$. 
Outgoing waves with  $k_n\in \mathbb{R}$ are found within the band $|\epsilon_0+n\omega|\!<\! 2\,J$, while  bound states outside the band have complex  $k_n=\mathrm{i}\kappa_n$ for $\epsilon_0+n\omega<-2\,J$ and $k_n=\mathrm{i}\kappa_n + \pi$ for $\epsilon_0+n\omega>2\,J$ with $\kappa_n \in \mathbb{R}$. Inserting the ansatz \eqref{eq:Fansatz} into \eqref{eq:H_floquet} we obtain $\delta_{n,0}+r_{n}=t_{n}=\frac{J'}{J}\xi_n $ as well as a set of coupled algebraic equations (see Appendix A)
\begin{align}
    \frac{\mu}{2}(t_{n+1}+t_{n-1})=\gamma_n t_n - \mathrm{i}\delta_{n0}\frac{J'^2}{J^2} v_0
\label{eq:t_n recursion}
\end{align}
where $v_n = 2J\sin(k_n)$ is the group velocity and
\begin{align}\label{gamma_n}
    \gamma_n = \left(1-\frac{J'^2}{J^2}\right) (\epsilon_0+n\omega) -\mu_0 +\im \frac{J'^2}{J^2}v_n.
\end{align}
The convergence condition
$\lim_{n\rightarrow \infty}t_{\pm n} \!=\!0$ ensures a unique solution, which we now find analytically in terms of continued fractions: 
Defining $c_{n}^\pm\!:=\!-\frac{\mu}{2} \frac{t_{\pm(n+ 1)}}{t_{\pm n}}  $ for $n\!\ge\! 0$ it follows from Eq.~(\ref{eq:t_n recursion}) that 
\begin{align}
c_n^\pm\! =\! \frac{-\mu^2}{4(\gamma_{\pm (n+ 1)}\! +\! c_{n+1}^\pm)}\! =\! \frac{-\mu^2}{ 4(\gamma_{\pm (n+ 1)}\!-\!\frac{\mu^2}{4(\gamma_{\pm(n+2)}+ {\dots}) })}
\label{eq:CF_rec}
\end{align}
where $\dots$ indicates a continued fraction obtained by straightforward iteration. Inserting $c_0^\pm$ from Eq.~(\ref{eq:CF_rec}) into Eq.~(\ref{eq:t_n recursion}) for $\!n=\!0$, the transmission coefficient becomes 
\begin{align}
t_0=\frac{\im v_0}{\gamma_0 +c_0^++c_0^-}\frac{J'^2}{J^2}  .
\label{eq:CF}
\end{align}
This formula now provides a central analytical result for the transmission of the general setup, which allows a deeper theoretical study and experimentally relevant predictions of perfect transport. The analytic expressions in Eqs.~(\ref{eq:CF_rec}) and (\ref{eq:CF}) can be evaluated to arbitrary precision  by iteratively evaluating the recursion  
starting from a large cutoff value of $n\!=\!M$, where $c_M^\pm$ is vanishingly small as long as $|\gamma_{\pm M}|\!\gg\! \mu^2$.  As shown in Fig.~\ref{fig:cplot_mu0} 
for $J'\!=\!0.15J$, $\mu\!=\!0.3J$, and $\epsilon_0\!=\!1.5J$ the resulting transmission amplitude $|t_0^2|$ has a rich structure
of resonances and minima. As discussed below the features are robust also for other 
parameter values where $J'$, $\mu$, $\mu_0$, $\epsilon_0$, and $\omega$ determine  the position and width of maxima and minima of $|t_0|^2$.
One striking feature is the near perfect transmission $| t_0^2|\! \to\! 1$ for $\omega+\epsilon_0\!> \!2J$ marked by the green dotted line in Fig.~\ref{fig:cplot_mu0}, i.e.~for transport through a  quantum dot level {\it outside the band}.  The locations of the maxima are analyzed in more detail in the following.

\begin{figure}[t]
    \centering
     \includegraphics[width=0.42\textwidth]{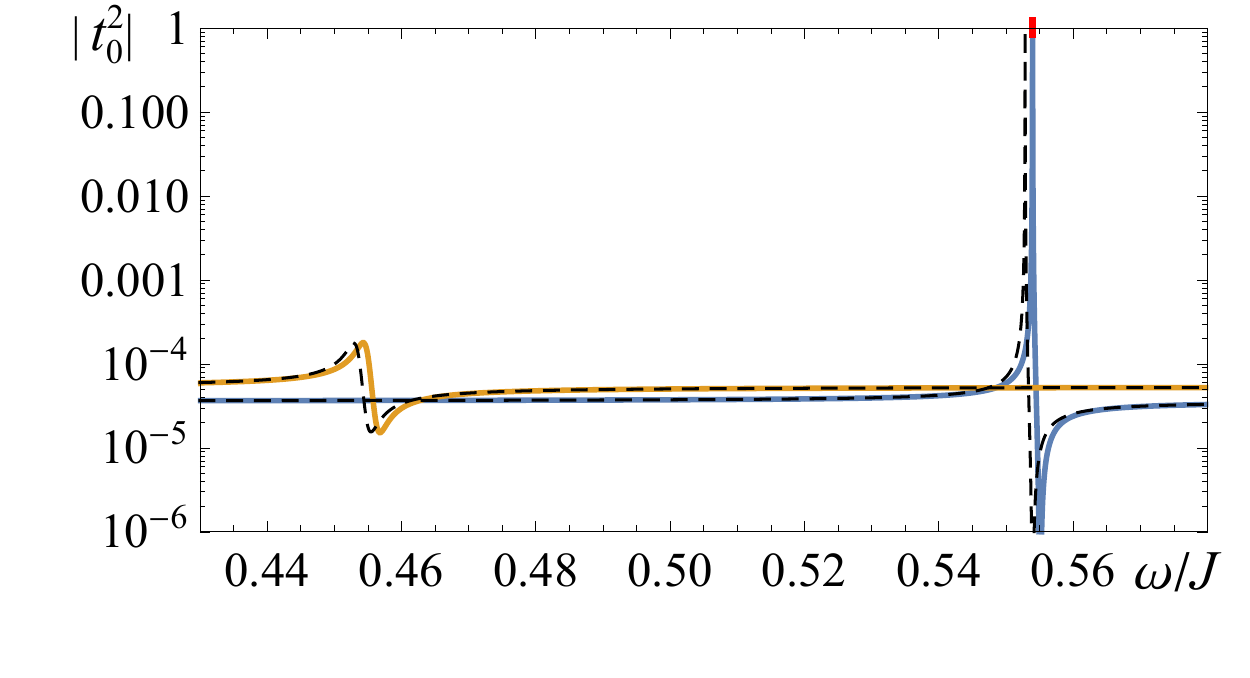}\vspace{-.5cm}
    \caption{Exact transmission $|t_0^2|$  for incoming energy $\epsilon_0\!=\!1.5\,J$ as a function of driving frequency $\omega$ for 
    $\mu\! =\! J'\!=\!0.05J$.  A dramatic difference is observed between the energy levels $\mu_0\!=\!0.195J$ (orange) and 
    $\mu_0\!=\!2.05J$ (blue) inside and outside the band, respectively.  The dashed curves use the lowest order
    approximations  $c^\pm_0\! \approx\! -\mu^2/(4 \gamma_{\pm1})$ in Eq.~(\ref{eq:CF}) and the prediction from Eq.~(\ref{res}) is marked in red on the top axis.\vspace{-.5cm}}
    \label{cut}
\end{figure}

{\it Small amplitudes $\mu\!\ll\! J$.} Let us first address the static case $\mu\!=\!0\!=\!c_0^\pm$.  According to Eqs.~(\ref{gamma_n}) and (\ref{eq:CF}) perfect transmission can be achieved by energy matching
$\epsilon_0\!=\! \mu_0/(1\!-\!\nicefrac{J'^2}{J^2})$.   However, we are interested in the generic experimental situation away from this special point, where the static transmission is suppressed $|t_0| \propto J'^2$ and now propose to use a small driving amplitudes $\mu$ for switching.  To lowest order 
in $\mu$ the continued fraction in Eq.~(\ref{eq:CF_rec}) is approximated by 
 $c^\pm_0\! \approx\! -\mu^2/(4 \gamma_{\pm1})$ in  Eq.~(\ref{eq:CF}).    Naively, this should only be a small correction of order $\mu^2$ to the static limit of $t_0$.  
 However, it is possible that $\gamma_1$  becomes very small of order $\mu^2$, which in turn allows a large change with  $c^+_0$ of the denominator in Eq.~(\ref{eq:CF}).
 In particular, in the limit of $\mu\to 0$, it follows that $\gamma_1\to 0$ gives the  position of the first resonance.
 Using Eq.~(\ref{gamma_n}) this requires $\im v_1 \!\in\! \mathbb{R} $ so that $\omega+\epsilon_0> 2J$ is outside the band. In this case, the resonance condition $\gamma_1\to 0$ evaluates to
 \begin{equation} \label{res}
     \omega \approx \frac{(J^2-J'^2)\mu_0-J'^2\sqrt{\mu_0^2-4J^2+8J'^2} }{J^2-2J'^2}-\epsilon_0, 
 \end{equation}
As can be seen in 
Fig.~\ref{fig:cplot_mu0} this prediction for the location of strong transmission agrees very well even for a sizable  value of $\mu=0.3J$ (dotted green line).
In Fig.~\ref{cut} we analyze the leading resonances on a logarithmic scale for very small values $\mu\!=\!J'\!=\!0.05J$ from Eq.~(\ref{eq:CF}), which agree well with the 
lowest order approximation  $c^\pm_0\! \approx\! -\mu^2/(4 \gamma_{\pm1})$.  Hence a very small control signal $\mu$ can lead to an enhancement by order of magnitudes at the right frequency.  We also see a dramatic difference between levels $\mu_0\!=\!2.05J$ 
outside the band and $\mu_0\!=\!1.95J$ inside the band, where the resonance transmission is much smaller.  The physical reason for the much lower values of $|t_0^2|$ is that a coupling to unbound Floquet components 
with $v_1\!\in\! \mathbb{R} $ effectively causes a large imaginary part in $\gamma_1$, leading to strong damping due to outgoing losses in excitations.  Non-resonant
unbound Floquet components with $v_{n \neq0}\!\in\! \mathbb{R} $ only contribute a very small imaginary part, which is for example the case for $n\!=\!-1$ and the parameters in Fig.~\ref{cut}, so the transmission is still close to unity as long as the leading velocity $\im v_1\!\in\! \mathbb{R} $ corresponds to a bound state. 
Moreover,  perfect transmission is always possible if there are no unbound Floquet components $\im v_{n}\!\in\! \mathbb{R}, \forall n\!\neq\! 0 $.
For the static case, quantum dot levels outside the band basically do not contribute to transport and receive little attention.  For the periodically driven system we now observe  the opposite behavior, that the transmission can be entirely dominated and controlled by those higher energy states.  

{\it Small tunneling $J'\!\ll\! J$.} For a more detailed theoretical analysis, we now consider small tunneling $J'\!\ll \!J$ to 
express the transmission amplitude from Eq.~\eqref{eq:CF}
\begin{align}\label{T}
    |t_0|^2\!=\!\frac{J'^4v_0^2}{(J'^2 v_0 \!+ \!J ^2 \mathrm{Im}[c_0^+ \!+\! c_0^-])^2\!+\!( J^2\mathrm{Re}[\gamma_0 \!+\!c_0^+\! +\!c_0^-])^2},
\end{align}
in terms of Bessel functions.  In the limit $J'\!\ll\! J$ the imaginary parts in Eq.~(\ref{gamma_n}) are small, so we first focus on the real part. 
In the following, we simplify notation by using a slightly  re-scaled frequency $\tilde{\omega}\!=\!(1\!-\!\nicefrac{J'^2}{J^2})\omega$ and energy $\tilde{\epsilon}_0\!=\!(1\!-\!\nicefrac{J'^2}{J^2})\epsilon_0$. 
Using the recurrence relation \cite{ContinuedFractions}  for  Bessel functions $2\nu\mathcal{J}_\nu\! =\!x(\mathcal{J}_{\nu+1}\!+\!\mathcal{J}_{\nu-1})$, we  define the ratio 
\begin{align} \label{bessel}
 a_\nu=x\frac{\mathcal{J}_{\nu +1}(x)}{\mathcal{J}_{\nu}(x)} \ \ {\rm obeyeing}\ \  a_\nu=\frac{x^2}{2(\nu+1)-a_{\nu+1}},
\end{align} 
which we recognize as the real part of the recursion relation in Eq.~(\ref{eq:CF_rec})  
in the limit of $J'\ll J$ if we identify
\begin{equation} \label{a_n}
{\rm Re} c_n^\pm =\mp\frac{\tilde \omega}{2} a_{n\pm\nu_0}, \ \     x=\frac{\mu}{\tilde{\omega}}, \ \  \nu_0=\frac{\tilde\epsilon_0-\mu_0}{\tilde{\omega}}.
\end{equation}
Note, that the contribution of order $\im J'^2 v_n/J^2$ in the denominator from Eq.~(\ref{gamma_n}) vanishes as $J'\to 0$ and has been omitted.  From Eqs.~(\ref{bessel}) and (\ref{a_n}) we
obtain 
\begin{equation}
     \mathrm{Re}(c_0^+ +c_0^-) = -\frac{\mu}{2}    \left( \frac{\mathcal{J}_{\nu_0+1}(x)}{\mathcal{J}_{\nu_0}(x)} -\frac{\mathcal{J}_{-\nu_0+1}(x)}{\mathcal{J}_{-\nu_0}(x)} \right).
     \label{eq:Rec0_0}
\end{equation}
According to Eq.~(\ref{T}) the transmission is large when this real part cancels ${\rm Re} \gamma_0$ in the denominator.  This
cancellation always occurs for $\nu_0 \in \mathbb{Z}$ due to the property 
of Bessel functions $\mathcal{J}_{-m}(x)=(-)^m\mathcal{J}_m(x)$ for integer $m\in \mathbb{Z}$ which simplifies Eq.~\eqref{eq:Rec0_0}
\begin{align}
     \mathrm{Re}(c_0^+ +c_0^-) = -\tilde{\epsilon}_0 +\mu_0=-\mathrm{Re}\gamma_0 \ \ {\rm if} \ \  \nu_0 \in \mathbb{Z}.
\end{align}
The condition $\nu_0 \in \mathbb{Z}$ for maxima becomes 
\begin{equation} \label{res2}
m\tilde \omega \approx |\mu_0-\tilde \epsilon_0|, \ \ m \in \mathbb{N},
\end{equation}
which agrees with Eq.~(\ref{res}) for $J'\!\to\! 0$ and $m\!=\!1$, but now predicts the locations both inside and outside the band of all maxima labelled by $m\!\geq\! 1$.

\begin{figure}[t]
    \centering
     \includegraphics[width=0.42\textwidth]{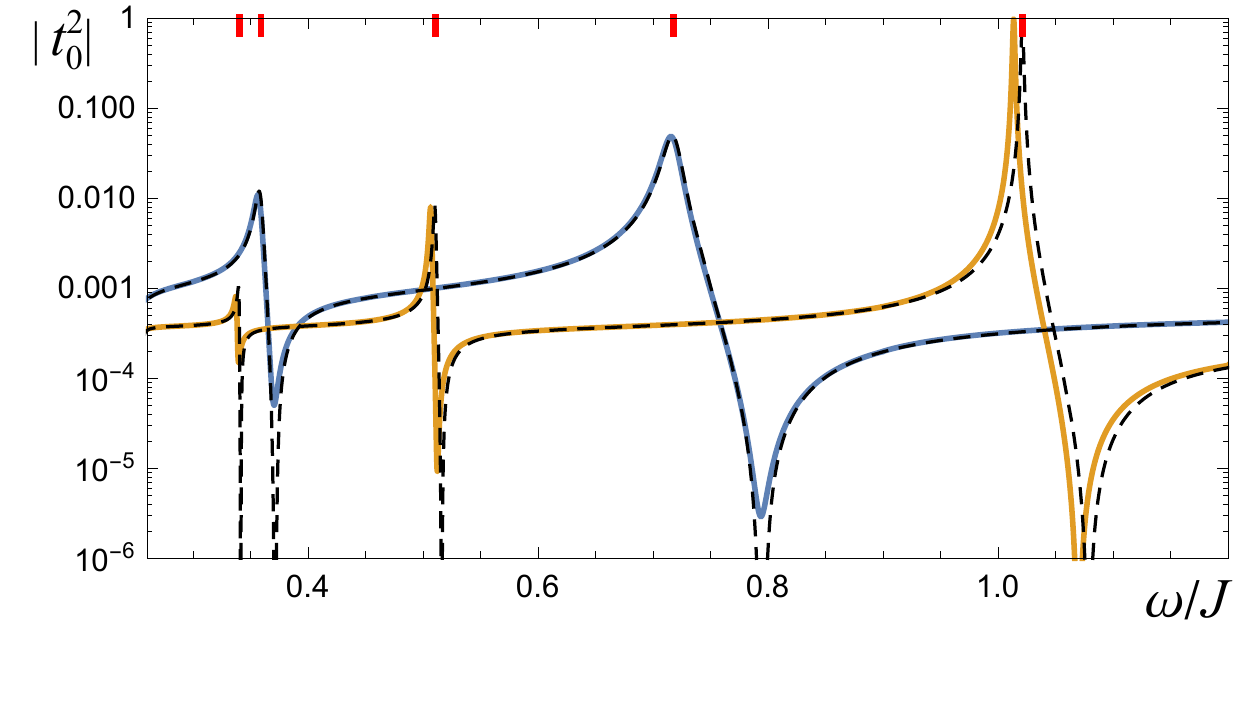}\vspace{-.5cm}
    \caption{Transmission $|t_0^2|$ as a function of driving frequency $\omega$ for  $\epsilon_0\!=\!1.1\,J$, 
    $\mu\! =\!0.5J$, $ J'\!=\!0.1J$ and two  different energy levels $\mu_0\!=\!2.1J$ (orange) and 
    $\mu_0\!=\!1.8J$ (blue).   The dashed curves  are the Bessel function approximations from Eqs.~(\ref{T}), (\ref{eq:Rec0_0}), and (\ref{impart}) and the maxima from Eq.~(\ref{res2}) are marked in red.\vspace{-.5cm}}
    \label{cut1}
\end{figure}
Finally, for a full closed form approximation we turn to the imaginary part in the deminator of Eq.~(\ref{T}).  As shown in Eqs.~(\ref{eq:c+}) and (\ref{eq:c-}) in Appendix B it can be approximated as 
\begin{align}
    \mathrm{Im}(c_0^++ c_0^-)  \approx \frac{J'^2}{J^2}{\sum_{n\neq 0}}^{'}  v_n\frac{\mathcal{J}_{{\rm sign}(n)(n+\nu_0)}^2(x)}{\mathcal{J}_{{\rm sign(n)}\nu_0}^2(x)},
    \label{impart}
\end{align}
where the primed sum is restricted to indices $n$ which correspond to  Floquet components inside the band $ v_n\! \in\! \mathbb{R} $, i.e.~$| \epsilon_0\!+\! n \omega |\!<\!2 J $.
As expected, the imaginary part is of order $\mathcal{O}(J'^2)$, but plays an important role for the absolute value of $| t_0^2|$ near resonances in Eq.~(\ref{T}).  Interestingly, if $\omega\!>\! 2J\!+\! | \epsilon_0| $ is large enough that no
Floquet components are inside the band, the imaginary part vanishes to all orders and perfect transmission can always be achieved by finding the zeros of the real part, i.e.~there exists a resonant value $\omega$  with $|t_0^2|=1$ for any given $\mu$ and  $J'\!<\!J$ if $|\mu_0-\epsilon_0| > 2J + |\epsilon_0|$,  which we confirmed numerically.

\begin{figure}[b]
    \centering
     \includegraphics[width=0.42\textwidth]{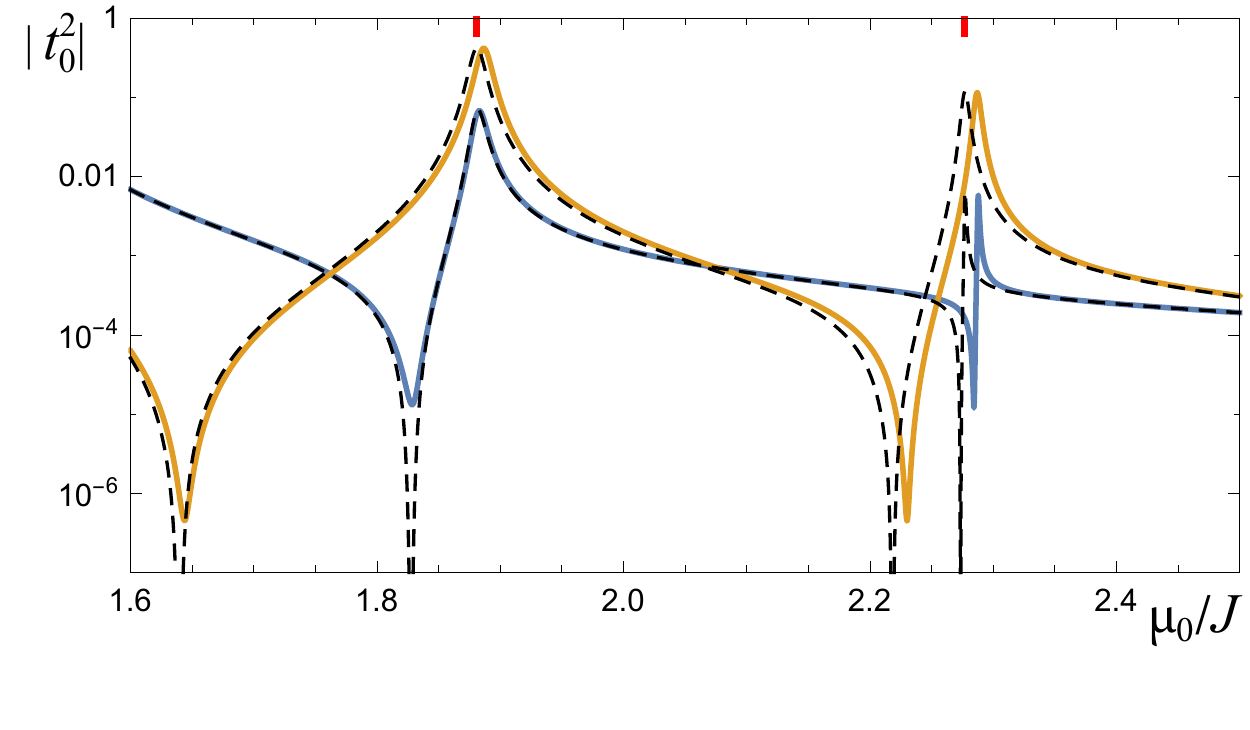}\vspace{-.5cm}
    \caption{Transmission $|t_0^2|$  for incoming energy $\epsilon_0\!=\!1.5\,J$ as a function of energy level $\mu_0$ for  $\omega\!=\!0.4J$, $ J'\!=\!0.1J$, and driving amplitudes 
    $\mu\! =\!0.3J$ (blue) and $\mu\! =\!0.7J$ (orange).    The dashed curves  are the Bessel function approximations from Eqs.~(\ref{T}), (\ref{eq:Rec0_0}), and (\ref{impart}) and the maxima from Eq.~(\ref{res2}) are marked in red.\vspace{-.3cm}}
    \label{cut2}
\end{figure}

\begin{figure}[t]
    \centering
     \includegraphics[width=0.49\textwidth]{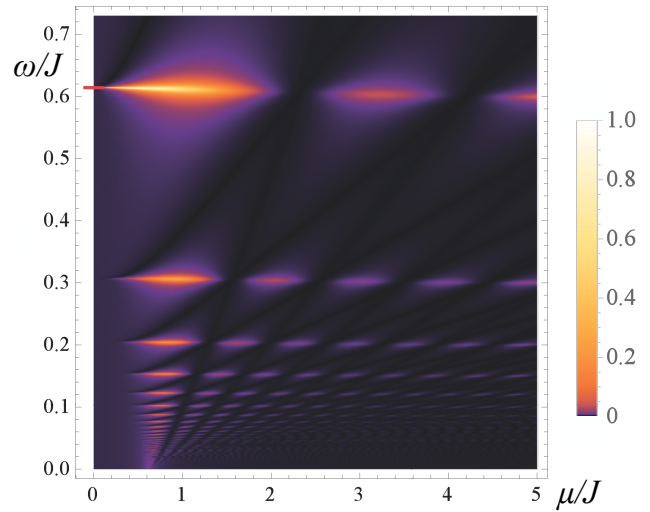}\vspace{-.3cm}
    \caption{Transmission $|t_0^2|$  for $\epsilon_0\!=\!1.5\,J$,  $\mu_0\!=\!2.1J$,  and $J'\! = \! 0.1J$ as a function of   $\omega$ and 
    $\mu$. The resonance position from Eq.~(\ref{res}) is marked red. \vspace{-.5cm}}
    \label{mu}
\end{figure}

A comparison of  Eqs.~(\ref{T}), (\ref{eq:Rec0_0}), and (\ref{impart}) with the exact result is shown in Fig.~\ref{cut1}  as a function of $\omega$ for  $\epsilon_0\!=\!1.1\,J$,     $\mu\! =\!0.5J$, $ J'\!=\!0.1J$, and two  different energy levels $\mu_0\!=\!2.1J$ (orange) and 
    $\mu_0\!=\!1.8J$ (blue).
Notably, the $m\!=\!1$ transmission resonance for $\mu_0\!=\!2.1J$ at $\omega\! \!\sim\!\!  J$ is outside the band and shows near perfect transmission, while all other maxima are at least an order of magnitude smaller.  Good 
agreement with the Bessel function approximations is observed,
except near the minima.  This is because the denominator in Eq.~(\ref{T}) diverges at the zeros of $\mathcal{J}_{\pm \nu_0}$
according to Eqs.~(\ref{eq:Rec0_0}) and (\ref{impart}), so the expansion breaks down. 
  Alternatively, we can also look at the behavior as a function of energy level $\mu_0$ in Fig.~\ref{cut2} for $\omega\!=\!0.4J$, $\epsilon_0\!=\!1.5J$, and two different driving amplitudes $\mu\! =\!0.3J$ (blue) and $\mu\!=\!0.7J$ (orange) in good agreement with  the closed form approximations.  
  For the larger amplitude  the maximum is broadened but the position is largely unchanged.
In fact, the predicted resonance positions in Eqs.~(\ref{res}) and (\ref{res2}) are independent of driving amplitude, so it is worthwhile to consider if this holds for larger values of $\mu$.
As shown in Fig.~\ref{mu} only small corrections to the resonant frequency are observed with increasing $\mu$, but the  
transmission undulates in agreement with the behavior of Bessel-functions. 

{\it Discussion.} It should be noted that approximations with Bessel functions are quite common in the Floquet description, where an effective Hamiltonian can 
be derived using a high frequency expansion \cite{eckardt2005,Eckardt_2015,pelster, Bukov04032015, wang, PhysRevX.4.031027}.  However in the derivation above, the Bessel functions now appear from an analysis of the recursion relation, which does
not require high frequencies, as can also be seen in the good agreement for  $\omega \!<\! J$ in the plots.  We would also like to point out that the limits $J'\! \ll\! J$
or $\mu\! \ll\! J$ discussed above
do not correspond to perturbation theory or Fermi's Golden Rule.    The                 approximations are
useful for a closed form analysis of the extrema, but of course  the full analytic solution in Eqs.~(\ref{eq:CF_rec}) and (\ref{eq:CF}) should be used for the most accurate predictions.  One notable property of the resonance structure is the observation of maxima for frequencies that are smaller than the energy gap in Eq.~(\ref{res2}), i.e.~a subharmonic effect which clearly cannot be due to excitations with single photons.  Instead, we understand the origin of 
the transmission maxima in terms of tunneling via dynamically created Floquet states with shifted virtual levels of the quantum dot.

 In conclusion, we have analyzed the transmission of particle transport through a periodically driven quantum dot. Our main finding is the possibility of switching to perfect transmission resonances with just a small oscillating control signal,  opening a great opportunity for designing switching devices with high tunability and accuracy. 
Essential for these resonances is the appearance of dynamically created Floquet levels from coupling to higher energy states outside the band which are not 
accessible in static scenarios. 
An analytic formula is obtained, describing the transmission amplitude in wide parameter regimes using continued fractions. This result allows for easy and computationally cheap calculations of the transmission on the one hand and accurate predictions of the resonance conditions on the other hand. In particular, we derive a closed analytic form of the transmission amplitude in the limit of weak tunneling to the quantum dot using Bessel-functions, which is not limited by a high frequency approximation. 

As for the physical implementation, the generic model in Eq.~(\ref{model}) does not make any reference to the type of quantum dot or leads, so 
molecules, tunneling tips, clusters, or semi-conductors can be used with an oscillating potential or alternatively with high frequency vibrations.  In fact, the model is also applicable to experimental transport setups in magnonics \cite{PhysRev.58.1098,10.1063/1.3499236,Shen_2018}, photonic waveguides \cite{waveguide, GARANOVICH20121,PhysRevResearch.3.013260}, or ultracold gases in optical lattices \cite{Bloch2005}.

\begin{acknowledgments}
We thank G.~Lefkidis for inspiring discussions.
Financial support was provided by the Deutsche Forschungsgemeinschaft (DFG, German Research Foundation) via the Collaborative Research Center SFB/TR185 (Project No. 277625399).
\end{acknowledgments}

\bibliography{bibliography.bib}

\begin{thebibliography}{64}%
\makeatletter
\providecommand \@ifxundefined [1]{%
 \@ifx{#1\undefined}
}%
\providecommand \@ifnum [1]{%
 \ifnum #1\expandafter \@firstoftwo
 \else \expandafter \@secondoftwo
 \fi
}%
\providecommand \@ifx [1]{%
 \ifx #1\expandafter \@firstoftwo
 \else \expandafter \@secondoftwo
 \fi
}%
\providecommand \natexlab [1]{#1}%
\providecommand \enquote  [1]{``#1''}%
\providecommand \bibnamefont  [1]{#1}%
\providecommand \bibfnamefont [1]{#1}%
\providecommand \citenamefont [1]{#1}%
\providecommand \href@noop [0]{\@secondoftwo}%
\providecommand \href [0]{\begingroup \@sanitize@url \@href}%
\providecommand \@href[1]{\@@startlink{#1}\@@href}%
\providecommand \@@href[1]{\endgroup#1\@@endlink}%
\providecommand \@sanitize@url [0]{\catcode `\\12\catcode `\$12\catcode
  `\&12\catcode `\#12\catcode `\^12\catcode `\_12\catcode `\%12\relax}%
\providecommand \@@startlink[1]{}%
\providecommand \@@endlink[0]{}%
\providecommand \url  [0]{\begingroup\@sanitize@url \@url }%
\providecommand \@url [1]{\endgroup\@href {#1}{\urlprefix }}%
\providecommand \urlprefix  [0]{URL }%
\providecommand \Eprint [0]{\href }%
\providecommand \doibase [0]{https://doi.org/}%
\providecommand \selectlanguage [0]{\@gobble}%
\providecommand \bibinfo  [0]{\@secondoftwo}%
\providecommand \bibfield  [0]{\@secondoftwo}%
\providecommand \translation [1]{[#1]}%
\providecommand \BibitemOpen [0]{}%
\providecommand \bibitemStop [0]{}%
\providecommand \bibitemNoStop [0]{.\EOS\space}%
\providecommand \EOS [0]{\spacefactor3000\relax}%
\providecommand \BibitemShut  [1]{\csname bibitem#1\endcsname}%
\let\auto@bib@innerbib\@empty
\bibitem [{\citenamefont {Aviram}\ and\ \citenamefont
  {Ratner}(1974)}]{Aviram1974}%
  \BibitemOpen
  \bibfield  {author} {\bibinfo {author} {\bibfnamefont {A.}~\bibnamefont
  {Aviram}}\ and\ \bibinfo {author} {\bibfnamefont {M.~A.}\ \bibnamefont
  {Ratner}},\ }\bibfield  {title} {\bibinfo {title} {Molecular rectifiers},\
  }\href {https://doi.org/10.1016/0009-2614(74)85031-1} {\bibfield  {journal}
  {\bibinfo  {journal} {Chemical Physics Letters}\ }\textbf {\bibinfo {volume}
  {29}},\ \bibinfo {pages} {277} (\bibinfo {year} {1974})}\BibitemShut
  {NoStop}%
\bibitem [{\citenamefont {Reed}\ \emph {et~al.}(1997)\citenamefont {Reed},
  \citenamefont {Zhou}, \citenamefont {Muller}, \citenamefont {Burgin},\ and\
  \citenamefont {Tour}}]{Reed1997}%
  \BibitemOpen
  \bibfield  {author} {\bibinfo {author} {\bibfnamefont {M.~A.}\ \bibnamefont
  {Reed}}, \bibinfo {author} {\bibfnamefont {C.}~\bibnamefont {Zhou}}, \bibinfo
  {author} {\bibfnamefont {C.~J.}\ \bibnamefont {Muller}}, \bibinfo {author}
  {\bibfnamefont {T.~P.}\ \bibnamefont {Burgin}},\ and\ \bibinfo {author}
  {\bibfnamefont {J.~M.}\ \bibnamefont {Tour}},\ }\bibfield  {title} {\bibinfo
  {title} {Conductance of a molecular junction},\ }\href
  {https://doi.org/10.1126/science.278.5336.252} {\bibfield  {journal}
  {\bibinfo  {journal} {Science}\ }\textbf {\bibinfo {volume} {278}},\ \bibinfo
  {pages} {252} (\bibinfo {year} {1997})}\BibitemShut {NoStop}%
\bibitem [{\citenamefont {Cui}\ \emph {et~al.}(2001)\citenamefont {Cui},
  \citenamefont {Primak}, \citenamefont {Zarate}, \citenamefont {Tomfohr},
  \citenamefont {Sankey}, \citenamefont {Moore}, \citenamefont {Moore},
  \citenamefont {Gust}, \citenamefont {Harris},\ and\ \citenamefont
  {Lindsay}}]{Cui2001}%
  \BibitemOpen
  \bibfield  {author} {\bibinfo {author} {\bibfnamefont {X.~D.}\ \bibnamefont
  {Cui}}, \bibinfo {author} {\bibfnamefont {A.}~\bibnamefont {Primak}},
  \bibinfo {author} {\bibfnamefont {X.}~\bibnamefont {Zarate}}, \bibinfo
  {author} {\bibfnamefont {J.}~\bibnamefont {Tomfohr}}, \bibinfo {author}
  {\bibfnamefont {O.~F.}\ \bibnamefont {Sankey}}, \bibinfo {author}
  {\bibfnamefont {A.~L.}\ \bibnamefont {Moore}}, \bibinfo {author}
  {\bibfnamefont {T.~A.}\ \bibnamefont {Moore}}, \bibinfo {author}
  {\bibfnamefont {D.}~\bibnamefont {Gust}}, \bibinfo {author} {\bibfnamefont
  {G.}~\bibnamefont {Harris}},\ and\ \bibinfo {author} {\bibfnamefont {S.~M.}\
  \bibnamefont {Lindsay}},\ }\bibfield  {title} {\bibinfo {title} {Reproducible
  measurement of single-molecule conductivity},\ }\href
  {https://doi.org/10.1126/science.1064354} {\bibfield  {journal} {\bibinfo
  {journal} {Science}\ }\textbf {\bibinfo {volume} {294}},\ \bibinfo {pages}
  {571} (\bibinfo {year} {2001})}\BibitemShut {NoStop}%
\bibitem [{\citenamefont {Fagas}\ \emph {et~al.}(2001)\citenamefont {Fagas},
  \citenamefont {Cuniberti},\ and\ \citenamefont
  {Richter}}]{PhysRevB.63.045416}%
  \BibitemOpen
  \bibfield  {author} {\bibinfo {author} {\bibfnamefont {G.}~\bibnamefont
  {Fagas}}, \bibinfo {author} {\bibfnamefont {G.}~\bibnamefont {Cuniberti}},\
  and\ \bibinfo {author} {\bibfnamefont {K.}~\bibnamefont {Richter}},\
  }\bibfield  {title} {\bibinfo {title} {Electron transport in
  nanotube--molecular-wire hybrids},\ }\href
  {https://doi.org/10.1103/PhysRevB.63.045416} {\bibfield  {journal} {\bibinfo
  {journal} {Phys. Rev. B}\ }\textbf {\bibinfo {volume} {63}},\ \bibinfo
  {pages} {045416} (\bibinfo {year} {2001})}\BibitemShut {NoStop}%
\bibitem [{\citenamefont {Gutierrez}\ \emph {et~al.}(2002)\citenamefont
  {Gutierrez}, \citenamefont {Fagas}, \citenamefont {Cuniberti}, \citenamefont
  {Grossmann}, \citenamefont {Schmidt},\ and\ \citenamefont
  {Richter}}]{PhysRevB.65.113410}%
  \BibitemOpen
  \bibfield  {author} {\bibinfo {author} {\bibfnamefont {R.}~\bibnamefont
  {Gutierrez}}, \bibinfo {author} {\bibfnamefont {G.}~\bibnamefont {Fagas}},
  \bibinfo {author} {\bibfnamefont {G.}~\bibnamefont {Cuniberti}}, \bibinfo
  {author} {\bibfnamefont {F.}~\bibnamefont {Grossmann}}, \bibinfo {author}
  {\bibfnamefont {R.}~\bibnamefont {Schmidt}},\ and\ \bibinfo {author}
  {\bibfnamefont {K.}~\bibnamefont {Richter}},\ }\bibfield  {title} {\bibinfo
  {title} {Theory of an all-carbon molecular switch},\ }\href
  {https://doi.org/10.1103/PhysRevB.65.113410} {\bibfield  {journal} {\bibinfo
  {journal} {Phys. Rev. B}\ }\textbf {\bibinfo {volume} {65}},\ \bibinfo
  {pages} {113410} (\bibinfo {year} {2002})}\BibitemShut {NoStop}%
\bibitem [{\citenamefont {Galperin}\ \emph {et~al.}(2008)\citenamefont
  {Galperin}, \citenamefont {Ratner}, \citenamefont {Nitzan},\ and\
  \citenamefont {Troisi}}]{Galperin2008}%
  \BibitemOpen
  \bibfield  {author} {\bibinfo {author} {\bibfnamefont {M.}~\bibnamefont
  {Galperin}}, \bibinfo {author} {\bibfnamefont {M.~A.}\ \bibnamefont
  {Ratner}}, \bibinfo {author} {\bibfnamefont {A.}~\bibnamefont {Nitzan}},\
  and\ \bibinfo {author} {\bibfnamefont {A.}~\bibnamefont {Troisi}},\
  }\bibfield  {title} {\bibinfo {title} {Nuclear coupling and polarization in
  molecular transport junctions: Beyond tunneling to function},\ }\href
  {https://doi.org/10.1126/science.1146556} {\bibfield  {journal} {\bibinfo
  {journal} {Science}\ }\textbf {\bibinfo {volume} {319}},\ \bibinfo {pages}
  {1056} (\bibinfo {year} {2008})}\BibitemShut {NoStop}%
\bibitem [{\citenamefont {Shen}\ \emph {et~al.}(2010)\citenamefont {Shen},
  \citenamefont {Zeng}, \citenamefont {Yang}, \citenamefont {Zhang},
  \citenamefont {Wang},\ and\ \citenamefont {Feng}}]{Shen2010}%
  \BibitemOpen
  \bibfield  {author} {\bibinfo {author} {\bibfnamefont {L.}~\bibnamefont
  {Shen}}, \bibinfo {author} {\bibfnamefont {M.}~\bibnamefont {Zeng}}, \bibinfo
  {author} {\bibfnamefont {S.-W.}\ \bibnamefont {Yang}}, \bibinfo {author}
  {\bibfnamefont {C.}~\bibnamefont {Zhang}}, \bibinfo {author} {\bibfnamefont
  {X.}~\bibnamefont {Wang}},\ and\ \bibinfo {author} {\bibfnamefont
  {Y.}~\bibnamefont {Feng}},\ }\bibfield  {title} {\bibinfo {title} {Electron
  transport properties of atomic carbon nanowires between graphene
  electrodes},\ }\href {https://doi.org/10.1021/ja909531c} {\bibfield
  {journal} {\bibinfo  {journal} {Journal of the American Chemical Society}\
  }\textbf {\bibinfo {volume} {132}},\ \bibinfo {pages} {11481} (\bibinfo
  {year} {2010})},\ \bibinfo {note} {pMID: 20677763}\BibitemShut {NoStop}%
\bibitem [{\citenamefont {Garrigues}\ \emph {et~al.}(2016)\citenamefont
  {Garrigues}, \citenamefont {Yuan}, \citenamefont {Wang}, \citenamefont
  {Mucciolo}, \citenamefont {Thompon}, \citenamefont {del Barco},\ and\
  \citenamefont {Nijhuis}}]{Garrigues2016}%
  \BibitemOpen
  \bibfield  {author} {\bibinfo {author} {\bibfnamefont {A.~R.}\ \bibnamefont
  {Garrigues}}, \bibinfo {author} {\bibfnamefont {L.}~\bibnamefont {Yuan}},
  \bibinfo {author} {\bibfnamefont {L.}~\bibnamefont {Wang}}, \bibinfo {author}
  {\bibfnamefont {E.~R.}\ \bibnamefont {Mucciolo}}, \bibinfo {author}
  {\bibfnamefont {D.}~\bibnamefont {Thompon}}, \bibinfo {author} {\bibfnamefont
  {E.}~\bibnamefont {del Barco}},\ and\ \bibinfo {author} {\bibfnamefont
  {C.~A.}\ \bibnamefont {Nijhuis}},\ }\bibfield  {title} {\bibinfo {title} {A
  single-level tunnel model to account for electrical transport through single
  molecule- and self-assembled monolayer-based junctions},\ }\href
  {https://doi.org/10.1038/srep26517} {\bibfield  {journal} {\bibinfo
  {journal} {Scientific Reports}\ }\textbf {\bibinfo {volume} {6}},\ \bibinfo
  {pages} {26517} (\bibinfo {year} {2016})}\BibitemShut {NoStop}%
\bibitem [{\citenamefont {Song}\ \emph {et~al.}(2020)\citenamefont {Song},
  \citenamefont {Han}, \citenamefont {Yu},\ and\ \citenamefont
  {Hu}}]{SONG2020110514}%
  \BibitemOpen
  \bibfield  {author} {\bibinfo {author} {\bibfnamefont {X.}~\bibnamefont
  {Song}}, \bibinfo {author} {\bibfnamefont {B.}~\bibnamefont {Han}}, \bibinfo
  {author} {\bibfnamefont {X.}~\bibnamefont {Yu}},\ and\ \bibinfo {author}
  {\bibfnamefont {W.}~\bibnamefont {Hu}},\ }\bibfield  {title} {\bibinfo
  {title} {The analysis of charge transport mechanism in molecular junctions
  based on current-voltage characteristics},\ }\href
  {https://doi.org/https://doi.org/10.1016/j.chemphys.2019.110514} {\bibfield
  {journal} {\bibinfo  {journal} {Chemical Physics}\ }\textbf {\bibinfo
  {volume} {528}},\ \bibinfo {pages} {110514} (\bibinfo {year}
  {2020})}\BibitemShut {NoStop}%
\bibitem [{\citenamefont {Werner}\ \emph {et~al.}(2024)\citenamefont {Werner},
  \citenamefont {\ifmmode~\check{Z}\else \v{Z}\fi{}itko},\ and\ \citenamefont
  {Arrigoni}}]{PhysRevB.109.075156}%
  \BibitemOpen
  \bibfield  {author} {\bibinfo {author} {\bibfnamefont {D.}~\bibnamefont
  {Werner}}, \bibinfo {author} {\bibfnamefont {R.}~\bibnamefont
  {\ifmmode~\check{Z}\else \v{Z}\fi{}itko}},\ and\ \bibinfo {author}
  {\bibfnamefont {E.}~\bibnamefont {Arrigoni}},\ }\bibfield  {title} {\bibinfo
  {title} {Auxiliary master equation approach to the anderson-holstein impurity
  problem out of equilibrium},\ }\href
  {https://doi.org/10.1103/PhysRevB.109.075156} {\bibfield  {journal} {\bibinfo
   {journal} {Phys. Rev. B}\ }\textbf {\bibinfo {volume} {109}},\ \bibinfo
  {pages} {075156} (\bibinfo {year} {2024})}\BibitemShut {NoStop}%
\bibitem [{\citenamefont {Yang}\ \emph {et~al.}(2023)\citenamefont {Yang},
  \citenamefont {Yang}, \citenamefont {Guo}, \citenamefont {Feng},\ and\
  \citenamefont {Guo}}]{Yang2023}%
  \BibitemOpen
  \bibfield  {author} {\bibinfo {author} {\bibfnamefont {C.}~\bibnamefont
  {Yang}}, \bibinfo {author} {\bibfnamefont {C.}~\bibnamefont {Yang}}, \bibinfo
  {author} {\bibfnamefont {Y.}~\bibnamefont {Guo}}, \bibinfo {author}
  {\bibfnamefont {J.}~\bibnamefont {Feng}},\ and\ \bibinfo {author}
  {\bibfnamefont {X.}~\bibnamefont {Guo}},\ }\bibfield  {title} {\bibinfo
  {title} {Graphene--molecule--graphene single-molecule junctions to detect
  electronic reactions at the molecular scale},\ }\href
  {https://doi.org/10.1038/s41596-023-00822-x} {\bibfield  {journal} {\bibinfo
  {journal} {Nature Protocols}\ }\textbf {\bibinfo {volume} {18}},\ \bibinfo
  {pages} {1958} (\bibinfo {year} {2023})}\BibitemShut {NoStop}%
\bibitem [{\citenamefont {Nitzan}\ and\ \citenamefont
  {Ratner}(2003)}]{Nitzan2003}%
  \BibitemOpen
  \bibfield  {author} {\bibinfo {author} {\bibfnamefont {A.}~\bibnamefont
  {Nitzan}}\ and\ \bibinfo {author} {\bibfnamefont {M.~A.}\ \bibnamefont
  {Ratner}},\ }\bibfield  {title} {\bibinfo {title} {Electron transport in
  molecular wire junctions},\ }\href {https://doi.org/10.1126/science.1081572}
  {\bibfield  {journal} {\bibinfo  {journal} {Science}\ }\textbf {\bibinfo
  {volume} {300}},\ \bibinfo {pages} {1384} (\bibinfo {year}
  {2003})}\BibitemShut {NoStop}%
\bibitem [{\citenamefont {Metzger}(2015)}]{Metzger2015}%
  \BibitemOpen
  \bibfield  {author} {\bibinfo {author} {\bibfnamefont {R.~M.}\ \bibnamefont
  {Metzger}},\ }\bibfield  {title} {\bibinfo {title} {Unimolecular
  electronics},\ }\href {https://doi.org/10.1021/cr500459d} {\bibfield
  {journal} {\bibinfo  {journal} {Chemical Reviews}\ }\textbf {\bibinfo
  {volume} {115}},\ \bibinfo {pages} {5056} (\bibinfo {year} {2015})},\
  \bibinfo {note} {pMID: 25950274}\BibitemShut {NoStop}%
\bibitem [{\citenamefont {Xiang}\ \emph {et~al.}(2016)\citenamefont {Xiang},
  \citenamefont {Wang}, \citenamefont {Jia}, \citenamefont {Lee},\ and\
  \citenamefont {Guo}}]{Xiang2016}%
  \BibitemOpen
  \bibfield  {author} {\bibinfo {author} {\bibfnamefont {D.}~\bibnamefont
  {Xiang}}, \bibinfo {author} {\bibfnamefont {X.}~\bibnamefont {Wang}},
  \bibinfo {author} {\bibfnamefont {C.}~\bibnamefont {Jia}}, \bibinfo {author}
  {\bibfnamefont {T.}~\bibnamefont {Lee}},\ and\ \bibinfo {author}
  {\bibfnamefont {X.}~\bibnamefont {Guo}},\ }\bibfield  {title} {\bibinfo
  {title} {Molecular-scale electronics: From concept to function},\ }\href
  {https://doi.org/10.1021/acs.chemrev.5b00680} {\bibfield  {journal} {\bibinfo
   {journal} {Chemical Reviews}\ }\textbf {\bibinfo {volume} {116}},\ \bibinfo
  {pages} {4318} (\bibinfo {year} {2016})},\ \bibinfo {note} {pMID:
  26979510}\BibitemShut {NoStop}%
\bibitem [{\citenamefont {Han}\ \emph {et~al.}(2024)\citenamefont {Han},
  \citenamefont {Liang}, \citenamefont {Razdolski}, \citenamefont {Bai},
  \citenamefont {Li},\ and\ \citenamefont {Lei}}]{HAN2024100517}%
  \BibitemOpen
  \bibfield  {author} {\bibinfo {author} {\bibfnamefont {S.}~\bibnamefont
  {Han}}, \bibinfo {author} {\bibfnamefont {X.}~\bibnamefont {Liang}}, \bibinfo
  {author} {\bibfnamefont {I.}~\bibnamefont {Razdolski}}, \bibinfo {author}
  {\bibfnamefont {Y.}~\bibnamefont {Bai}}, \bibinfo {author} {\bibfnamefont
  {H.}~\bibnamefont {Li}},\ and\ \bibinfo {author} {\bibfnamefont
  {D.}~\bibnamefont {Lei}},\ }\bibfield  {title} {\bibinfo {title} {Optical and
  charge transport characteristics of photoswitching plasmonic molecular
  systems},\ }\href
  {https://doi.org/https://doi.org/10.1016/j.pquantelec.2024.100517} {\bibfield
   {journal} {\bibinfo  {journal} {Progress in Quantum Electronics}\ }\textbf
  {\bibinfo {volume} {95}},\ \bibinfo {pages} {100517} (\bibinfo {year}
  {2024})}\BibitemShut {NoStop}%
\bibitem [{\citenamefont {Platero}\ and\ \citenamefont
  {Aguado}(2004)}]{PLATERO20041}%
  \BibitemOpen
  \bibfield  {author} {\bibinfo {author} {\bibfnamefont {G.}~\bibnamefont
  {Platero}}\ and\ \bibinfo {author} {\bibfnamefont {R.}~\bibnamefont
  {Aguado}},\ }\bibfield  {title} {\bibinfo {title} {Photon-assisted transport
  in semiconductor nanostructures},\ }\href
  {https://doi.org/https://doi.org/10.1016/j.physrep.2004.01.004} {\bibfield
  {journal} {\bibinfo  {journal} {Physics Reports}\ }\textbf {\bibinfo {volume}
  {395}},\ \bibinfo {pages} {1} (\bibinfo {year} {2004})}\BibitemShut {NoStop}%
\bibitem [{\citenamefont {Yuan}\ \emph {et~al.}(2014)\citenamefont {Yuan},
  \citenamefont {Zhu}, \citenamefont {Zhao}, \citenamefont {Guo}, \citenamefont
  {Yan}, \citenamefont {Sun},\ and\ \citenamefont {Ji}}]{Yuan2014}%
  \BibitemOpen
  \bibfield  {author} {\bibinfo {author} {\bibfnamefont {R.~Y.}\ \bibnamefont
  {Yuan}}, \bibinfo {author} {\bibfnamefont {G.~B.}\ \bibnamefont {Zhu}},
  \bibinfo {author} {\bibfnamefont {X.}~\bibnamefont {Zhao}}, \bibinfo {author}
  {\bibfnamefont {Y.}~\bibnamefont {Guo}}, \bibinfo {author} {\bibfnamefont
  {H.}~\bibnamefont {Yan}}, \bibinfo {author} {\bibfnamefont {Q.}~\bibnamefont
  {Sun}},\ and\ \bibinfo {author} {\bibfnamefont {A.~C.}\ \bibnamefont {Ji}},\
  }\bibfield  {title} {\bibinfo {title} {Coulomb interaction effects on the
  terahertz photon-assisted tunneling through an {I}n{A}s quantum dot},\ }\href
  {https://doi.org/10.1103/PHYSREVB.89.195301/FIGURES/4/MEDIUM} {\bibfield
  {journal} {\bibinfo  {journal} {Physical Review B - Condensed Matter and
  Materials Physics}\ }\textbf {\bibinfo {volume} {89}},\ \bibinfo {pages}
  {195301} (\bibinfo {year} {2014})}\BibitemShut {NoStop}%
\bibitem [{\citenamefont {Sun}\ and\ \citenamefont {Lin}(1997)}]{Sun1997}%
  \BibitemOpen
  \bibfield  {author} {\bibinfo {author} {\bibfnamefont {Q.~F.}\ \bibnamefont
  {Sun}}\ and\ \bibinfo {author} {\bibfnamefont {T.~H.}\ \bibnamefont {Lin}},\
  }\bibfield  {title} {\bibinfo {title} {Influence of microwave fields on the
  electron tunneling through a quantum dot},\ }\href
  {https://doi.org/10.1103/PhysRevB.56.3591} {\bibfield  {journal} {\bibinfo
  {journal} {Physical Review B}\ }\textbf {\bibinfo {volume} {56}},\ \bibinfo
  {pages} {3591} (\bibinfo {year} {1997})}\BibitemShut {NoStop}%
\bibitem [{\citenamefont {Jauho}\ \emph {et~al.}(1994)\citenamefont {Jauho},
  \citenamefont {Wingreen},\ and\ \citenamefont {Meir}}]{Jauho1994}%
  \BibitemOpen
  \bibfield  {author} {\bibinfo {author} {\bibfnamefont {A.~P.}\ \bibnamefont
  {Jauho}}, \bibinfo {author} {\bibfnamefont {N.~S.}\ \bibnamefont
  {Wingreen}},\ and\ \bibinfo {author} {\bibfnamefont {Y.}~\bibnamefont
  {Meir}},\ }\bibfield  {title} {\bibinfo {title} {Time-dependent transport in
  interacting and noninteracting resonant-tunneling systems},\ }\href
  {https://doi.org/10.1103/PhysRevB.50.5528} {\bibfield  {journal} {\bibinfo
  {journal} {Physical Review B}\ }\textbf {\bibinfo {volume} {50}},\ \bibinfo
  {pages} {5528} (\bibinfo {year} {1994})}\BibitemShut {NoStop}%
\bibitem [{\citenamefont {Qin}\ \emph {et~al.}(2001)\citenamefont {Qin},
  \citenamefont {Holleitner}, \citenamefont {Blick},\ and\ \citenamefont
  {Eberl}}]{Qin2001}%
  \BibitemOpen
  \bibfield  {author} {\bibinfo {author} {\bibfnamefont {H.}~\bibnamefont
  {Qin}}, \bibinfo {author} {\bibfnamefont {A.~W.}\ \bibnamefont {Holleitner}},
  \bibinfo {author} {\bibfnamefont {R.~H.}\ \bibnamefont {Blick}},\ and\
  \bibinfo {author} {\bibfnamefont {K.}~\bibnamefont {Eberl}},\ }\bibfield
  {title} {\bibinfo {title} {Coherent superposition of photon- and
  phonon-assisted tunneling in coupled quantum dots},\ }\href
  {https://doi.org/10.1103/PhysRevB.64.241302} {\bibfield  {journal} {\bibinfo
  {journal} {Physical Review B}\ }\textbf {\bibinfo {volume} {64}},\ \bibinfo
  {pages} {241302} (\bibinfo {year} {2001})}\BibitemShut {NoStop}%
\bibitem [{\citenamefont {Yang}\ \emph {et~al.}(2014)\citenamefont {Yang},
  \citenamefont {Liu}, \citenamefont {Wang},\ and\ \citenamefont
  {He}}]{Yang2014}%
  \BibitemOpen
  \bibfield  {author} {\bibinfo {author} {\bibfnamefont {K.~H.}\ \bibnamefont
  {Yang}}, \bibinfo {author} {\bibfnamefont {B.~Y.}\ \bibnamefont {Liu}},
  \bibinfo {author} {\bibfnamefont {H.~Y.}\ \bibnamefont {Wang}},\ and\
  \bibinfo {author} {\bibfnamefont {X.}~\bibnamefont {He}},\ }\bibfield
  {title} {\bibinfo {title} {Phonon-assisted tunneling and two-channel {K}ondo
  effect in a vibrating molecular dot coupled to {L}uttinger liquid leads},\
  }\href {https://doi.org/10.1016/J.SSC.2013.10.025} {\bibfield  {journal}
  {\bibinfo  {journal} {Solid State Communications}\ }\textbf {\bibinfo
  {volume} {178}},\ \bibinfo {pages} {50} (\bibinfo {year} {2014})}\BibitemShut
  {NoStop}%
\bibitem [{\citenamefont {Shafranjuk}(2008)}]{Shafranjuk2008}%
  \BibitemOpen
  \bibfield  {author} {\bibinfo {author} {\bibfnamefont {S.~E.}\ \bibnamefont
  {Shafranjuk}},\ }\bibfield  {title} {\bibinfo {title} {Probing the intrinsic
  state of a one-dimensional quantum well with photon-assisted tunneling},\
  }\href {https://doi.org/10.1103/PHYSREVB.78.235115/FIGURES/5/MEDIUM}
  {\bibfield  {journal} {\bibinfo  {journal} {Physical Review B - Condensed
  Matter and Materials Physics}\ }\textbf {\bibinfo {volume} {78}},\ \bibinfo
  {pages} {235115} (\bibinfo {year} {2008})}\BibitemShut {NoStop}%
\bibitem [{\citenamefont {Yang}\ \emph {et~al.}(2017)\citenamefont {Yang},
  \citenamefont {Qin}, \citenamefont {Wang},\ and\ \citenamefont
  {Liu}}]{Yang2017}%
  \BibitemOpen
  \bibfield  {author} {\bibinfo {author} {\bibfnamefont {K.~H.}\ \bibnamefont
  {Yang}}, \bibinfo {author} {\bibfnamefont {C.~D.}\ \bibnamefont {Qin}},
  \bibinfo {author} {\bibfnamefont {H.~Y.}\ \bibnamefont {Wang}},\ and\
  \bibinfo {author} {\bibfnamefont {K.~D.}\ \bibnamefont {Liu}},\ }\bibfield
  {title} {\bibinfo {title} {Effects of {L}uttinger leads on the {AC}
  conductance of a quantum dot},\ }\href
  {https://doi.org/10.1016/J.PHYSLETA.2017.02.017} {\bibfield  {journal}
  {\bibinfo  {journal} {Physics Letters A}\ }\textbf {\bibinfo {volume}
  {381}},\ \bibinfo {pages} {1328} (\bibinfo {year} {2017})}\BibitemShut
  {NoStop}%
\bibitem [{\citenamefont {Drexler}\ \emph {et~al.}(1995)\citenamefont
  {Drexler}, \citenamefont {Scott}, \citenamefont {Allen}, \citenamefont
  {Campman},\ and\ \citenamefont {Gossard}}]{Drexler1995}%
  \BibitemOpen
  \bibfield  {author} {\bibinfo {author} {\bibfnamefont {H.}~\bibnamefont
  {Drexler}}, \bibinfo {author} {\bibfnamefont {J.~S.}\ \bibnamefont {Scott}},
  \bibinfo {author} {\bibfnamefont {S.~J.}\ \bibnamefont {Allen}}, \bibinfo
  {author} {\bibfnamefont {K.~L.}\ \bibnamefont {Campman}},\ and\ \bibinfo
  {author} {\bibfnamefont {A.~C.}\ \bibnamefont {Gossard}},\ }\bibfield
  {title} {\bibinfo {title} {Photon‐assisted tunneling in a resonant
  tunneling diode: Stimulated emission and absorption in the {TH}z range},\
  }\href {https://doi.org/10.1063/1.114794} {\bibfield  {journal} {\bibinfo
  {journal} {Applied Physics Letters}\ }\textbf {\bibinfo {volume} {67}},\
  \bibinfo {pages} {2816} (\bibinfo {year} {1995})}\BibitemShut {NoStop}%
\bibitem [{\citenamefont {Kouwenhoven}\ \emph
  {et~al.}(1994{\natexlab{a}})\citenamefont {Kouwenhoven}, \citenamefont
  {Jauhar}, \citenamefont {McCormick}, \citenamefont {Dixon}, \citenamefont
  {McEuen}, \citenamefont {Nazarov}, \citenamefont {van~der Vaart},\ and\
  \citenamefont {Foxon}}]{PhysRevB.50.2019}%
  \BibitemOpen
  \bibfield  {author} {\bibinfo {author} {\bibfnamefont {L.~P.}\ \bibnamefont
  {Kouwenhoven}}, \bibinfo {author} {\bibfnamefont {S.}~\bibnamefont {Jauhar}},
  \bibinfo {author} {\bibfnamefont {K.}~\bibnamefont {McCormick}}, \bibinfo
  {author} {\bibfnamefont {D.}~\bibnamefont {Dixon}}, \bibinfo {author}
  {\bibfnamefont {P.~L.}\ \bibnamefont {McEuen}}, \bibinfo {author}
  {\bibfnamefont {Y.~V.}\ \bibnamefont {Nazarov}}, \bibinfo {author}
  {\bibfnamefont {N.~C.}\ \bibnamefont {van~der Vaart}},\ and\ \bibinfo
  {author} {\bibfnamefont {C.~T.}\ \bibnamefont {Foxon}},\ }\bibfield  {title}
  {\bibinfo {title} {Photon-assisted tunneling through a quantum dot},\ }\href
  {https://doi.org/10.1103/PhysRevB.50.2019} {\bibfield  {journal} {\bibinfo
  {journal} {Phys. Rev. B}\ }\textbf {\bibinfo {volume} {50}},\ \bibinfo
  {pages} {2019} (\bibinfo {year} {1994}{\natexlab{a}})}\BibitemShut {NoStop}%
\bibitem [{\citenamefont {Kouwenhoven}\ \emph
  {et~al.}(1994{\natexlab{b}})\citenamefont {Kouwenhoven}, \citenamefont
  {Jauhar}, \citenamefont {Orenstein}, \citenamefont {McEuen}, \citenamefont
  {Nagamune}, \citenamefont {Motohisa},\ and\ \citenamefont
  {Sakaki}}]{Kouwenhoven1994}%
  \BibitemOpen
  \bibfield  {author} {\bibinfo {author} {\bibfnamefont {L.~P.}\ \bibnamefont
  {Kouwenhoven}}, \bibinfo {author} {\bibfnamefont {S.}~\bibnamefont {Jauhar}},
  \bibinfo {author} {\bibfnamefont {J.}~\bibnamefont {Orenstein}}, \bibinfo
  {author} {\bibfnamefont {P.~L.}\ \bibnamefont {McEuen}}, \bibinfo {author}
  {\bibfnamefont {Y.}~\bibnamefont {Nagamune}}, \bibinfo {author}
  {\bibfnamefont {J.}~\bibnamefont {Motohisa}},\ and\ \bibinfo {author}
  {\bibfnamefont {H.}~\bibnamefont {Sakaki}},\ }\bibfield  {title} {\bibinfo
  {title} {Observation of photon-assisted tunneling through a quantum dot},\
  }\href {https://doi.org/10.1103/PhysRevLett.73.3443} {\bibfield  {journal}
  {\bibinfo  {journal} {Physical Review Letters}\ }\textbf {\bibinfo {volume}
  {73}},\ \bibinfo {pages} {3443} (\bibinfo {year}
  {1994}{\natexlab{b}})}\BibitemShut {NoStop}%
\bibitem [{\citenamefont {Lee}\ and\ \citenamefont {Tse}(2019)}]{Lee2019}%
  \BibitemOpen
  \bibfield  {author} {\bibinfo {author} {\bibfnamefont {W.~R.}\ \bibnamefont
  {Lee}}\ and\ \bibinfo {author} {\bibfnamefont {W.~K.}\ \bibnamefont {Tse}},\
  }\bibfield  {title} {\bibinfo {title} {Photon-induced suppression of
  interlayer tunneling in van der {W}aals heterostructures},\ }\href
  {https://doi.org/10.1103/PHYSREVB.99.201403/SUPP_RESUB1.PDF} {\bibfield
  {journal} {\bibinfo  {journal} {Physical Review B}\ }\textbf {\bibinfo
  {volume} {99}},\ \bibinfo {pages} {201403} (\bibinfo {year}
  {2019})}\BibitemShut {NoStop}%
\bibitem [{\citenamefont {Kouwenhoven}\ \emph {et~al.}(1991)\citenamefont
  {Kouwenhoven}, \citenamefont {Johnson}, \citenamefont {van~der Vaart},
  \citenamefont {Harmans},\ and\ \citenamefont {Foxon}}]{Kouwenhoven1991}%
  \BibitemOpen
  \bibfield  {author} {\bibinfo {author} {\bibfnamefont {L.~P.}\ \bibnamefont
  {Kouwenhoven}}, \bibinfo {author} {\bibfnamefont {A.~T.}\ \bibnamefont
  {Johnson}}, \bibinfo {author} {\bibfnamefont {N.~C.}\ \bibnamefont {van~der
  Vaart}}, \bibinfo {author} {\bibfnamefont {C.~J. P.~M.}\ \bibnamefont
  {Harmans}},\ and\ \bibinfo {author} {\bibfnamefont {C.~T.}\ \bibnamefont
  {Foxon}},\ }\bibfield  {title} {\bibinfo {title} {Quantized current in a
  quantum-dot turnstile using oscillating tunnel barriers},\ }\href
  {https://doi.org/10.1103/PhysRevLett.67.1626} {\bibfield  {journal} {\bibinfo
   {journal} {Phys. Rev. Lett.}\ }\textbf {\bibinfo {volume} {67}},\ \bibinfo
  {pages} {1626} (\bibinfo {year} {1991})}\BibitemShut {NoStop}%
\bibitem [{\citenamefont {Blick}\ \emph {et~al.}(1995)\citenamefont {Blick},
  \citenamefont {Haug}, \citenamefont {van~der Weide}, \citenamefont {von
  Klitzing},\ and\ \citenamefont {Eberl}}]{Blick_1995}%
  \BibitemOpen
  \bibfield  {author} {\bibinfo {author} {\bibfnamefont {R.~H.}\ \bibnamefont
  {Blick}}, \bibinfo {author} {\bibfnamefont {R.~J.}\ \bibnamefont {Haug}},
  \bibinfo {author} {\bibfnamefont {D.~W.}\ \bibnamefont {van~der Weide}},
  \bibinfo {author} {\bibfnamefont {K.}~\bibnamefont {von Klitzing}},\ and\
  \bibinfo {author} {\bibfnamefont {K.}~\bibnamefont {Eberl}},\ }\bibfield
  {title} {\bibinfo {title} {Photon‐assisted tunneling through a quantum dot
  at high microwave frequencies},\ }\href {https://doi.org/10.1063/1.114406}
  {\bibfield  {journal} {\bibinfo  {journal} {Applied Physics Letters}\
  }\textbf {\bibinfo {volume} {67}},\ \bibinfo {pages} {3924} (\bibinfo {year}
  {1995})}\BibitemShut {NoStop}%
\bibitem [{\citenamefont {Iñarrea}\ \emph {et~al.}(1997)\citenamefont
  {Iñarrea}, \citenamefont {Aguado},\ and\ \citenamefont
  {Platero}}]{Inarrea_1997}%
  \BibitemOpen
  \bibfield  {author} {\bibinfo {author} {\bibfnamefont {J.}~\bibnamefont
  {Iñarrea}}, \bibinfo {author} {\bibfnamefont {R.}~\bibnamefont {Aguado}},\
  and\ \bibinfo {author} {\bibfnamefont {G.}~\bibnamefont {Platero}},\
  }\bibfield  {title} {\bibinfo {title} {Electron-photon interaction in
  resonant tunneling diodes},\ }\href
  {https://doi.org/10.1209/epl/i1997-00481-1} {\bibfield  {journal} {\bibinfo
  {journal} {Europhysics Letters}\ }\textbf {\bibinfo {volume} {40}},\ \bibinfo
  {pages} {417} (\bibinfo {year} {1997})}\BibitemShut {NoStop}%
\bibitem [{\citenamefont {Oosterkamp}\ \emph {et~al.}(1997)\citenamefont
  {Oosterkamp}, \citenamefont {Kouwenhoven}, \citenamefont {Koolen},
  \citenamefont {van~der Vaart},\ and\ \citenamefont
  {Harmans}}]{PhysRevLett.78.1536}%
  \BibitemOpen
  \bibfield  {author} {\bibinfo {author} {\bibfnamefont {T.~H.}\ \bibnamefont
  {Oosterkamp}}, \bibinfo {author} {\bibfnamefont {L.~P.}\ \bibnamefont
  {Kouwenhoven}}, \bibinfo {author} {\bibfnamefont {A.~E.~A.}\ \bibnamefont
  {Koolen}}, \bibinfo {author} {\bibfnamefont {N.~C.}\ \bibnamefont {van~der
  Vaart}},\ and\ \bibinfo {author} {\bibfnamefont {C.~J. P.~M.}\ \bibnamefont
  {Harmans}},\ }\bibfield  {title} {\bibinfo {title} {Photon sidebands of the
  ground state and first excited state of a quantum dot},\ }\href
  {https://doi.org/10.1103/PhysRevLett.78.1536} {\bibfield  {journal} {\bibinfo
   {journal} {Phys. Rev. Lett.}\ }\textbf {\bibinfo {volume} {78}},\ \bibinfo
  {pages} {1536} (\bibinfo {year} {1997})}\BibitemShut {NoStop}%
\bibitem [{\citenamefont {Creffield}\ and\ \citenamefont
  {Platero}(2002)}]{PhysRevB.65.113304}%
  \BibitemOpen
  \bibfield  {author} {\bibinfo {author} {\bibfnamefont {C.~E.}\ \bibnamefont
  {Creffield}}\ and\ \bibinfo {author} {\bibfnamefont {G.}~\bibnamefont
  {Platero}},\ }\bibfield  {title} {\bibinfo {title} {{AC}-driven localization
  in a two-electron quantum dot molecule},\ }\href
  {https://doi.org/10.1103/PhysRevB.65.113304} {\bibfield  {journal} {\bibinfo
  {journal} {Phys. Rev. B}\ }\textbf {\bibinfo {volume} {65}},\ \bibinfo
  {pages} {113304} (\bibinfo {year} {2002})}\BibitemShut {NoStop}%
\bibitem [{\citenamefont {Shibata}\ \emph {et~al.}(2012)\citenamefont
  {Shibata}, \citenamefont {Umeno}, \citenamefont {Cha},\ and\ \citenamefont
  {Hirakawa}}]{Shibata2012}%
  \BibitemOpen
  \bibfield  {author} {\bibinfo {author} {\bibfnamefont {K.}~\bibnamefont
  {Shibata}}, \bibinfo {author} {\bibfnamefont {A.}~\bibnamefont {Umeno}},
  \bibinfo {author} {\bibfnamefont {K.~M.}\ \bibnamefont {Cha}},\ and\ \bibinfo
  {author} {\bibfnamefont {K.}~\bibnamefont {Hirakawa}},\ }\bibfield  {title}
  {\bibinfo {title} {Photon-assisted tunneling through self-assembled {I}n{A}s
  quantum dots in the terahertz frequency range},\ }\href
  {https://doi.org/10.1103/PHYSREVLETT.109.077401/FIGURES/5/MEDIUM} {\bibfield
  {journal} {\bibinfo  {journal} {Physical Review Letters}\ }\textbf {\bibinfo
  {volume} {109}},\ \bibinfo {pages} {077401} (\bibinfo {year}
  {2012})}\BibitemShut {NoStop}%
\bibitem [{\citenamefont {Ortega-Taberner}\ \emph {et~al.}(2023)\citenamefont
  {Ortega-Taberner}, \citenamefont {Jauho},\ and\ \citenamefont
  {Paaske}}]{PhysRevB.107.115165}%
  \BibitemOpen
  \bibfield  {author} {\bibinfo {author} {\bibfnamefont {C.}~\bibnamefont
  {Ortega-Taberner}}, \bibinfo {author} {\bibfnamefont {A.-P.}\ \bibnamefont
  {Jauho}},\ and\ \bibinfo {author} {\bibfnamefont {J.}~\bibnamefont
  {Paaske}},\ }\bibfield  {title} {\bibinfo {title} {Anomalous {J}osephson
  current through a driven double quantum dot},\ }\href
  {https://doi.org/10.1103/PhysRevB.107.115165} {\bibfield  {journal} {\bibinfo
   {journal} {Phys. Rev. B}\ }\textbf {\bibinfo {volume} {107}},\ \bibinfo
  {pages} {115165} (\bibinfo {year} {2023})}\BibitemShut {NoStop}%
\bibitem [{\citenamefont {Venitucci}\ \emph {et~al.}(2018)\citenamefont
  {Venitucci}, \citenamefont {Feinberg}, \citenamefont {M\'elin},\ and\
  \citenamefont {Dou\ifmmode~\mbox{\c{c}}\else
  \c{c}\fi{}ot}}]{PhysRevB.97.195423}%
  \BibitemOpen
  \bibfield  {author} {\bibinfo {author} {\bibfnamefont {B.}~\bibnamefont
  {Venitucci}}, \bibinfo {author} {\bibfnamefont {D.}~\bibnamefont {Feinberg}},
  \bibinfo {author} {\bibfnamefont {R.}~\bibnamefont {M\'elin}},\ and\ \bibinfo
  {author} {\bibfnamefont {B.}~\bibnamefont {Dou\ifmmode~\mbox{\c{c}}\else
  \c{c}\fi{}ot}},\ }\bibfield  {title} {\bibinfo {title} {Nonadiabatic
  {J}osephson current pumping by chiral microwave irradiation},\ }\href
  {https://doi.org/10.1103/PhysRevB.97.195423} {\bibfield  {journal} {\bibinfo
  {journal} {Phys. Rev. B}\ }\textbf {\bibinfo {volume} {97}},\ \bibinfo
  {pages} {195423} (\bibinfo {year} {2018})}\BibitemShut {NoStop}%
\bibitem [{\citenamefont {Mosallanejad}\ \emph {et~al.}(2024)\citenamefont
  {Mosallanejad}, \citenamefont {Wang},\ and\ \citenamefont
  {Dou}}]{10.1063/5.0184978}%
  \BibitemOpen
  \bibfield  {author} {\bibinfo {author} {\bibfnamefont {V.}~\bibnamefont
  {Mosallanejad}}, \bibinfo {author} {\bibfnamefont {Y.}~\bibnamefont {Wang}},\
  and\ \bibinfo {author} {\bibfnamefont {W.}~\bibnamefont {Dou}},\ }\bibfield
  {title} {\bibinfo {title} {Floquet non-equilibrium {G}reen’s function and
  {F}loquet quantum master equation for electronic transport: The role of
  electron–electron interactions and spin current with circular light},\
  }\href {https://doi.org/10.1063/5.0184978} {\bibfield  {journal} {\bibinfo
  {journal} {The Journal of Chemical Physics}\ }\textbf {\bibinfo {volume}
  {160}},\ \bibinfo {pages} {164102} (\bibinfo {year} {2024})}\BibitemShut
  {NoStop}%
\bibitem [{\citenamefont {Thuberg}\ \emph {et~al.}(2016)\citenamefont
  {Thuberg}, \citenamefont {Reyes},\ and\ \citenamefont
  {Eggert}}]{PhysRevB.93.180301}%
  \BibitemOpen
  \bibfield  {author} {\bibinfo {author} {\bibfnamefont {D.}~\bibnamefont
  {Thuberg}}, \bibinfo {author} {\bibfnamefont {S.~A.}\ \bibnamefont {Reyes}},\
  and\ \bibinfo {author} {\bibfnamefont {S.}~\bibnamefont {Eggert}},\
  }\bibfield  {title} {\bibinfo {title} {Quantum resonance catastrophe for
  conductance through a periodically driven barrier},\ }\href
  {https://doi.org/10.1103/PhysRevB.93.180301} {\bibfield  {journal} {\bibinfo
  {journal} {Phys. Rev. B}\ }\textbf {\bibinfo {volume} {93}},\ \bibinfo
  {pages} {180301} (\bibinfo {year} {2016})}\BibitemShut {NoStop}%
\bibitem [{\citenamefont {Reyes}\ \emph {et~al.}(2017)\citenamefont {Reyes},
  \citenamefont {Thuberg}, \citenamefont {Pérez}, \citenamefont {Dauer},\ and\
  \citenamefont {Eggert}}]{Reyes_2017}%
  \BibitemOpen
  \bibfield  {author} {\bibinfo {author} {\bibfnamefont {S.~A.}\ \bibnamefont
  {Reyes}}, \bibinfo {author} {\bibfnamefont {D.}~\bibnamefont {Thuberg}},
  \bibinfo {author} {\bibfnamefont {D.}~\bibnamefont {Pérez}}, \bibinfo
  {author} {\bibfnamefont {C.}~\bibnamefont {Dauer}},\ and\ \bibinfo {author}
  {\bibfnamefont {S.}~\bibnamefont {Eggert}},\ }\bibfield  {title} {\bibinfo
  {title} {Transport through an {AC}-driven impurity: Fano interference and
  bound states in the continuum},\ }\href
  {https://doi.org/10.1088/1367-2630/aa66fe} {\bibfield  {journal} {\bibinfo
  {journal} {New Journal of Physics}\ }\textbf {\bibinfo {volume} {19}},\
  \bibinfo {pages} {043029} (\bibinfo {year} {2017})}\BibitemShut {NoStop}%
\bibitem [{\citenamefont {Thuberg}\ \emph {et~al.}(2017)\citenamefont
  {Thuberg}, \citenamefont {Mu\~noz}, \citenamefont {Eggert},\ and\
  \citenamefont {Reyes}}]{Thuberg2017a}%
  \BibitemOpen
  \bibfield  {author} {\bibinfo {author} {\bibfnamefont {D.}~\bibnamefont
  {Thuberg}}, \bibinfo {author} {\bibfnamefont {E.}~\bibnamefont {Mu\~noz}},
  \bibinfo {author} {\bibfnamefont {S.}~\bibnamefont {Eggert}},\ and\ \bibinfo
  {author} {\bibfnamefont {S.~A.}\ \bibnamefont {Reyes}},\ }\bibfield  {title}
  {\bibinfo {title} {Perfect spin filter by periodic drive of a ferromagnetic
  quantum barrier},\ }\href {https://doi.org/10.1103/PhysRevLett.119.267701}
  {\bibfield  {journal} {\bibinfo  {journal} {Phys. Rev. Lett.}\ }\textbf
  {\bibinfo {volume} {119}},\ \bibinfo {pages} {267701} (\bibinfo {year}
  {2017})}\BibitemShut {NoStop}%
\bibitem [{\citenamefont {Agarwala}\ and\ \citenamefont
  {Sen}(2017)}]{PhysRevB.96.104309}%
  \BibitemOpen
  \bibfield  {author} {\bibinfo {author} {\bibfnamefont {A.}~\bibnamefont
  {Agarwala}}\ and\ \bibinfo {author} {\bibfnamefont {D.}~\bibnamefont {Sen}},\
  }\bibfield  {title} {\bibinfo {title} {Effects of local periodic driving on
  transport and generation of bound states},\ }\href
  {https://doi.org/10.1103/PhysRevB.96.104309} {\bibfield  {journal} {\bibinfo
  {journal} {Phys. Rev. B}\ }\textbf {\bibinfo {volume} {96}},\ \bibinfo
  {pages} {104309} (\bibinfo {year} {2017})}\BibitemShut {NoStop}%
\bibitem [{\citenamefont {H\"ubner}\ \emph {et~al.}(2022)\citenamefont
  {H\"ubner}, \citenamefont {Dauer}, \citenamefont {Eggert}, \citenamefont
  {Kollath},\ and\ \citenamefont {Sheikhan}}]{Huebner2021}%
  \BibitemOpen
  \bibfield  {author} {\bibinfo {author} {\bibfnamefont {F.}~\bibnamefont
  {H\"ubner}}, \bibinfo {author} {\bibfnamefont {C.}~\bibnamefont {Dauer}},
  \bibinfo {author} {\bibfnamefont {S.}~\bibnamefont {Eggert}}, \bibinfo
  {author} {\bibfnamefont {C.}~\bibnamefont {Kollath}},\ and\ \bibinfo {author}
  {\bibfnamefont {A.}~\bibnamefont {Sheikhan}},\ }\bibfield  {title} {\bibinfo
  {title} {Floquet-engineered pair and single particle filter in the {F}ermi
  {H}ubbard model},\ }\href@noop {} {\bibfield  {journal} {\bibinfo  {journal}
  {Phys. Rev. A}\ }\textbf {\bibinfo {volume} {106}},\ \bibinfo {pages}
  {043303} (\bibinfo {year} {2022})}\BibitemShut {NoStop}%
\bibitem [{\citenamefont {H\"ubner}\ \emph {et~al.}(2023)\citenamefont
  {H\"ubner}, \citenamefont {Dauer}, \citenamefont {Eggert}, \citenamefont
  {Kollath},\ and\ \citenamefont {Sheikhan}}]{PhysRevA.108.023307}%
  \BibitemOpen
  \bibfield  {author} {\bibinfo {author} {\bibfnamefont {F.}~\bibnamefont
  {H\"ubner}}, \bibinfo {author} {\bibfnamefont {C.}~\bibnamefont {Dauer}},
  \bibinfo {author} {\bibfnamefont {S.}~\bibnamefont {Eggert}}, \bibinfo
  {author} {\bibfnamefont {C.}~\bibnamefont {Kollath}},\ and\ \bibinfo {author}
  {\bibfnamefont {A.}~\bibnamefont {Sheikhan}},\ }\bibfield  {title} {\bibinfo
  {title} {Momentum-resolved {F}loquet-engineered pair and single-particle
  filter in the {F}ermi-{H}ubbard model},\ }\href
  {https://doi.org/10.1103/PhysRevA.108.023307} {\bibfield  {journal} {\bibinfo
   {journal} {Phys. Rev. A}\ }\textbf {\bibinfo {volume} {108}},\ \bibinfo
  {pages} {023307} (\bibinfo {year} {2023})}\BibitemShut {NoStop}%
\bibitem [{\citenamefont {Floquet}(1883)}]{Floquet}%
  \BibitemOpen
  \bibfield  {author} {\bibinfo {author} {\bibfnamefont {G.}~\bibnamefont
  {Floquet}},\ }\bibfield  {title} {\bibinfo {title} {Sur les \'equations
  diff\'erentielles lin\'eaires \`a coefficients p\'eriodiques},\ }\href
  {https://doi.org/10.24033/asens.220} {\bibfield  {journal} {\bibinfo
  {journal} {Annales scientifiques de l'\'Ecole Normale Sup\'erieure}\ }\textbf
  {\bibinfo {volume} {2e s{\'e}rie, 12}},\ \bibinfo {pages} {47} (\bibinfo
  {year} {1883})}\BibitemShut {NoStop}%
\bibitem [{\citenamefont {Shirley}(1965)}]{shirley}%
  \BibitemOpen
  \bibfield  {author} {\bibinfo {author} {\bibfnamefont {J.~H.}\ \bibnamefont
  {Shirley}},\ }\bibfield  {title} {\bibinfo {title} {Solution of the
  {S}chr\"odinger equation with a {H}amiltonian periodic in time},\ }\href
  {https://doi.org/10.1103/PhysRev.138.B979} {\bibfield  {journal} {\bibinfo
  {journal} {Phys. Rev.}\ }\textbf {\bibinfo {volume} {138}},\ \bibinfo {pages}
  {B979} (\bibinfo {year} {1965})}\BibitemShut {NoStop}%
\bibitem [{\citenamefont {Eckardt}(2017)}]{eckardt}%
  \BibitemOpen
  \bibfield  {author} {\bibinfo {author} {\bibfnamefont {A.}~\bibnamefont
  {Eckardt}},\ }\bibfield  {title} {\bibinfo {title} {Colloquium: Atomic
  quantum gases in periodically driven optical lattices},\ }\href
  {https://doi.org/10.1103/RevModPhys.89.011004} {\bibfield  {journal}
  {\bibinfo  {journal} {Rev. Mod. Phys.}\ }\textbf {\bibinfo {volume} {89}},\
  \bibinfo {pages} {011004} (\bibinfo {year} {2017})}\BibitemShut {NoStop}%
\bibitem [{\citenamefont {Holthaus}(2015)}]{Holthaus}%
  \BibitemOpen
  \bibfield  {author} {\bibinfo {author} {\bibfnamefont {M.}~\bibnamefont
  {Holthaus}},\ }\bibfield  {title} {\bibinfo {title} {Floquet engineering with
  quasienergy bands of periodically driven optical lattices},\ }\href
  {https://doi.org/10.1088/0953-4075/49/1/013001} {\bibfield  {journal}
  {\bibinfo  {journal} {Journal of Physics B: Atomic, Molecular and Optical
  Physics}\ }\textbf {\bibinfo {volume} {49}},\ \bibinfo {pages} {013001}
  (\bibinfo {year} {2015})}\BibitemShut {NoStop}%
\bibitem [{\citenamefont {Meirav}\ \emph {et~al.}(1990)\citenamefont {Meirav},
  \citenamefont {Kastner},\ and\ \citenamefont {Wind}}]{wind}%
  \BibitemOpen
  \bibfield  {author} {\bibinfo {author} {\bibfnamefont {U.}~\bibnamefont
  {Meirav}}, \bibinfo {author} {\bibfnamefont {M.~A.}\ \bibnamefont
  {Kastner}},\ and\ \bibinfo {author} {\bibfnamefont {S.~J.}\ \bibnamefont
  {Wind}},\ }\bibfield  {title} {\bibinfo {title} {Single-electron charging and
  periodic conductance resonances in {G}a{A}s nanostructures},\ }\href
  {https://doi.org/10.1103/PhysRevLett.65.771} {\bibfield  {journal} {\bibinfo
  {journal} {Phys. Rev. Lett.}\ }\textbf {\bibinfo {volume} {65}},\ \bibinfo
  {pages} {771} (\bibinfo {year} {1990})}\BibitemShut {NoStop}%
\bibitem [{\citenamefont {Fulton}\ and\ \citenamefont {Dolan}(1987)}]{dolan}%
  \BibitemOpen
  \bibfield  {author} {\bibinfo {author} {\bibfnamefont {T.~A.}\ \bibnamefont
  {Fulton}}\ and\ \bibinfo {author} {\bibfnamefont {G.~J.}\ \bibnamefont
  {Dolan}},\ }\bibfield  {title} {\bibinfo {title} {Observation of
  single-electron charging effects in small tunnel junctions},\ }\href
  {https://doi.org/10.1103/PhysRevLett.59.109} {\bibfield  {journal} {\bibinfo
  {journal} {Phys. Rev. Lett.}\ }\textbf {\bibinfo {volume} {59}},\ \bibinfo
  {pages} {109} (\bibinfo {year} {1987})}\BibitemShut {NoStop}%
\bibitem [{\citenamefont {Meir}\ and\ \citenamefont
  {Wingreen}(1992)}]{wingreen}%
  \BibitemOpen
  \bibfield  {author} {\bibinfo {author} {\bibfnamefont {Y.}~\bibnamefont
  {Meir}}\ and\ \bibinfo {author} {\bibfnamefont {N.~S.}\ \bibnamefont
  {Wingreen}},\ }\bibfield  {title} {\bibinfo {title} {Landauer formula for the
  current through an interacting electron region},\ }\href
  {https://doi.org/10.1103/PhysRevLett.68.2512} {\bibfield  {journal} {\bibinfo
   {journal} {Phys. Rev. Lett.}\ }\textbf {\bibinfo {volume} {68}},\ \bibinfo
  {pages} {2512} (\bibinfo {year} {1992})}\BibitemShut {NoStop}%
\bibitem [{\citenamefont {Kastner}(1992)}]{RevModPhys.64.849}%
  \BibitemOpen
  \bibfield  {author} {\bibinfo {author} {\bibfnamefont {M.~A.}\ \bibnamefont
  {Kastner}},\ }\bibfield  {title} {\bibinfo {title} {The single-electron
  transistor},\ }\href {https://doi.org/10.1103/RevModPhys.64.849} {\bibfield
  {journal} {\bibinfo  {journal} {Rev. Mod. Phys.}\ }\textbf {\bibinfo {volume}
  {64}},\ \bibinfo {pages} {849} (\bibinfo {year} {1992})}\BibitemShut
  {NoStop}%
\bibitem [{\citenamefont {Holstein}\ and\ \citenamefont
  {Primakoff}(1940)}]{PhysRev.58.1098}%
  \BibitemOpen
  \bibfield  {author} {\bibinfo {author} {\bibfnamefont {T.}~\bibnamefont
  {Holstein}}\ and\ \bibinfo {author} {\bibfnamefont {H.}~\bibnamefont
  {Primakoff}},\ }\bibfield  {title} {\bibinfo {title} {Field dependence of the
  intrinsic domain magnetization of a ferromagnet},\ }\href
  {https://doi.org/10.1103/PhysRev.58.1098} {\bibfield  {journal} {\bibinfo
  {journal} {Phys. Rev.}\ }\textbf {\bibinfo {volume} {58}},\ \bibinfo {pages}
  {1098} (\bibinfo {year} {1940})}\BibitemShut {NoStop}%
\bibitem [{\citenamefont {Petrov}\ and\ \citenamefont
  {Ostrovsky}(2010)}]{10.1063/1.3499236}%
  \BibitemOpen
  \bibfield  {author} {\bibinfo {author} {\bibfnamefont {E.~G.}\ \bibnamefont
  {Petrov}}\ and\ \bibinfo {author} {\bibfnamefont {V.}~\bibnamefont
  {Ostrovsky}},\ }\bibfield  {title} {\bibinfo {title} {Single-magnon tunneling
  through a ferromagnetic nanochain},\ }\href
  {https://doi.org/10.1063/1.3499236} {\bibfield  {journal} {\bibinfo
  {journal} {Low Temperature Physics}\ }\textbf {\bibinfo {volume} {36}},\
  \bibinfo {pages} {761} (\bibinfo {year} {2010})}\BibitemShut {NoStop}%
\bibitem [{\citenamefont {Shen}(2018)}]{Shen_2018}%
  \BibitemOpen
  \bibfield  {author} {\bibinfo {author} {\bibfnamefont {K.}~\bibnamefont
  {Shen}},\ }\bibfield  {title} {\bibinfo {title} {Finite temperature magnon
  spectra in yttrium iron garnet from a mean field approach in a tight-binding
  model},\ }\href {https://doi.org/10.1088/1367-2630/aab951} {\bibfield
  {journal} {\bibinfo  {journal} {New Journal of Physics}\ }\textbf {\bibinfo
  {volume} {20}},\ \bibinfo {pages} {043025} (\bibinfo {year}
  {2018})}\BibitemShut {NoStop}%
\bibitem [{\citenamefont {Wu}\ \emph {et~al.}(2021)\citenamefont {Wu},
  \citenamefont {Song}, \citenamefont {Gao}, \citenamefont {Chen},
  \citenamefont {Zhu},\ and\ \citenamefont {Li}}]{waveguide}%
  \BibitemOpen
  \bibfield  {author} {\bibinfo {author} {\bibfnamefont {S.}~\bibnamefont
  {Wu}}, \bibinfo {author} {\bibfnamefont {W.}~\bibnamefont {Song}}, \bibinfo
  {author} {\bibfnamefont {S.}~\bibnamefont {Gao}}, \bibinfo {author}
  {\bibfnamefont {Y.}~\bibnamefont {Chen}}, \bibinfo {author} {\bibfnamefont
  {S.}~\bibnamefont {Zhu}},\ and\ \bibinfo {author} {\bibfnamefont
  {T.}~\bibnamefont {Li}},\ }\bibfield  {title} {\bibinfo {title} {Floquet
  $\ensuremath{\pi}$ mode engineering in non-hermitian waveguide lattices},\
  }\href {https://doi.org/10.1103/PhysRevResearch.3.023211} {\bibfield
  {journal} {\bibinfo  {journal} {Phys. Rev. Res.}\ }\textbf {\bibinfo {volume}
  {3}},\ \bibinfo {pages} {023211} (\bibinfo {year} {2021})}\BibitemShut
  {NoStop}%
\bibitem [{\citenamefont {Garanovich}\ \emph {et~al.}(2012)\citenamefont
  {Garanovich}, \citenamefont {Longhi}, \citenamefont {Sukhorukov},\ and\
  \citenamefont {Kivshar}}]{GARANOVICH20121}%
  \BibitemOpen
  \bibfield  {author} {\bibinfo {author} {\bibfnamefont {I.~L.}\ \bibnamefont
  {Garanovich}}, \bibinfo {author} {\bibfnamefont {S.}~\bibnamefont {Longhi}},
  \bibinfo {author} {\bibfnamefont {A.~A.}\ \bibnamefont {Sukhorukov}},\ and\
  \bibinfo {author} {\bibfnamefont {Y.~S.}\ \bibnamefont {Kivshar}},\
  }\bibfield  {title} {\bibinfo {title} {Light propagation and localization in
  modulated photonic lattices and waveguides},\ }\href
  {https://doi.org/https://doi.org/10.1016/j.physrep.2012.03.005} {\bibfield
  {journal} {\bibinfo  {journal} {Physics Reports}\ }\textbf {\bibinfo {volume}
  {518}},\ \bibinfo {pages} {1} (\bibinfo {year} {2012})}\BibitemShut {NoStop}%
\bibitem [{\citenamefont {Fedorova}\ \emph {et~al.}(2021)\citenamefont
  {Fedorova}, \citenamefont {Dauer}, \citenamefont {Sidorenko}, \citenamefont
  {Eggert}, \citenamefont {Kroha},\ and\ \citenamefont
  {Linden}}]{PhysRevResearch.3.013260}%
  \BibitemOpen
  \bibfield  {author} {\bibinfo {author} {\bibfnamefont {Z.}~\bibnamefont
  {Fedorova}}, \bibinfo {author} {\bibfnamefont {C.}~\bibnamefont {Dauer}},
  \bibinfo {author} {\bibfnamefont {A.}~\bibnamefont {Sidorenko}}, \bibinfo
  {author} {\bibfnamefont {S.}~\bibnamefont {Eggert}}, \bibinfo {author}
  {\bibfnamefont {J.}~\bibnamefont {Kroha}},\ and\ \bibinfo {author}
  {\bibfnamefont {S.}~\bibnamefont {Linden}},\ }\bibfield  {title} {\bibinfo
  {title} {Dissipation engineered directional filter for quantum ratchets},\
  }\href {https://doi.org/10.1103/PhysRevResearch.3.013260} {\bibfield
  {journal} {\bibinfo  {journal} {Phys. Rev. Res.}\ }\textbf {\bibinfo {volume}
  {3}},\ \bibinfo {pages} {013260} (\bibinfo {year} {2021})}\BibitemShut
  {NoStop}%
\bibitem [{\citenamefont {Bloch}(2005)}]{Bloch2005}%
  \BibitemOpen
  \bibfield  {author} {\bibinfo {author} {\bibfnamefont {I.}~\bibnamefont
  {Bloch}},\ }\bibfield  {title} {\bibinfo {title} {Ultracold quantum gases in
  optical lattices},\ }\href {https://doi.org/10.1038/nphys138} {\bibfield
  {journal} {\bibinfo  {journal} {Nature Physics}\ }\textbf {\bibinfo {volume}
  {1}},\ \bibinfo {pages} {23} (\bibinfo {year} {2005})}\BibitemShut {NoStop}%
\bibitem [{\citenamefont {Jones}\ and\ \citenamefont
  {Thron}(1980)}]{ContinuedFractions}%
  \BibitemOpen
  \bibfield  {author} {\bibinfo {author} {\bibfnamefont {W.~B.}\ \bibnamefont
  {Jones}}\ and\ \bibinfo {author} {\bibfnamefont {W.~J.}\ \bibnamefont
  {Thron}},\ }\href@noop {} {\emph {\bibinfo {title} {Continued fractions
  analytic theory and applications}}},\ Encyclopedia of mathematics and its
  applications 11 : Section: Analysis\ (\bibinfo  {publisher}
  {Addison-Wesley},\ \bibinfo {address} {Reading, Mass. [u.a},\ \bibinfo {year}
  {1980})\BibitemShut {NoStop}%
\bibitem [{\citenamefont {Eckardt}\ \emph {et~al.}(2005)\citenamefont
  {Eckardt}, \citenamefont {Weiss},\ and\ \citenamefont
  {Holthaus}}]{eckardt2005}%
  \BibitemOpen
  \bibfield  {author} {\bibinfo {author} {\bibfnamefont {A.}~\bibnamefont
  {Eckardt}}, \bibinfo {author} {\bibfnamefont {C.}~\bibnamefont {Weiss}},\
  and\ \bibinfo {author} {\bibfnamefont {M.}~\bibnamefont {Holthaus}},\
  }\bibfield  {title} {\bibinfo {title} {Superfluid-insulator transition in a
  periodically driven optical lattice},\ }\href
  {https://doi.org/10.1103/PhysRevLett.95.260404} {\bibfield  {journal}
  {\bibinfo  {journal} {Phys. Rev. Lett.}\ }\textbf {\bibinfo {volume} {95}},\
  \bibinfo {pages} {260404} (\bibinfo {year} {2005})}\BibitemShut {NoStop}%
\bibitem [{\citenamefont {Eckardt}\ and\ \citenamefont
  {Anisimovas}(2015)}]{Eckardt_2015}%
  \BibitemOpen
  \bibfield  {author} {\bibinfo {author} {\bibfnamefont {A.}~\bibnamefont
  {Eckardt}}\ and\ \bibinfo {author} {\bibfnamefont {E.}~\bibnamefont
  {Anisimovas}},\ }\bibfield  {title} {\bibinfo {title} {High-frequency
  approximation for periodically driven quantum systems from a {F}loquet-space
  perspective},\ }\href {https://doi.org/10.1088/1367-2630/17/9/093039}
  {\bibfield  {journal} {\bibinfo  {journal} {New Journal of Physics}\ }\textbf
  {\bibinfo {volume} {17}},\ \bibinfo {pages} {093039} (\bibinfo {year}
  {2015})}\BibitemShut {NoStop}%
\bibitem [{\citenamefont {Wang}\ \emph {et~al.}(2014)\citenamefont {Wang},
  \citenamefont {Zhang}, \citenamefont {Santos}, \citenamefont {Eggert},\ and\
  \citenamefont {Pelster}}]{pelster}%
  \BibitemOpen
  \bibfield  {author} {\bibinfo {author} {\bibfnamefont {T.}~\bibnamefont
  {Wang}}, \bibinfo {author} {\bibfnamefont {X.-F.}\ \bibnamefont {Zhang}},
  \bibinfo {author} {\bibfnamefont {F.~E. A.~d.}\ \bibnamefont {Santos}},
  \bibinfo {author} {\bibfnamefont {S.}~\bibnamefont {Eggert}},\ and\ \bibinfo
  {author} {\bibfnamefont {A.}~\bibnamefont {Pelster}},\ }\bibfield  {title}
  {\bibinfo {title} {Tuning the quantum phase transition of bosons in optical
  lattices via periodic modulation of the $s$-wave scattering length},\ }\href
  {https://doi.org/10.1103/PhysRevA.90.013633} {\bibfield  {journal} {\bibinfo
  {journal} {Phys. Rev. A}\ }\textbf {\bibinfo {volume} {90}},\ \bibinfo
  {pages} {013633} (\bibinfo {year} {2014})}\BibitemShut {NoStop}%
\bibitem [{\citenamefont {Bukov}\ \emph {et~al.}(2015)\citenamefont {Bukov},
  \citenamefont {D'Alessio},\ and\ \citenamefont
  {Polkovnikov}}]{Bukov04032015}%
  \BibitemOpen
  \bibfield  {author} {\bibinfo {author} {\bibfnamefont {M.}~\bibnamefont
  {Bukov}}, \bibinfo {author} {\bibfnamefont {L.}~\bibnamefont {D'Alessio}},\
  and\ \bibinfo {author} {\bibfnamefont {A.}~\bibnamefont {Polkovnikov}},\
  }\bibfield  {title} {\bibinfo {title} {Universal high-frequency behavior of
  periodically driven systems: from dynamical stabilization to {F}loquet
  engineering},\ }\href {https://doi.org/10.1080/00018732.2015.1055918}
  {\bibfield  {journal} {\bibinfo  {journal} {Advances in Physics}\ }\textbf
  {\bibinfo {volume} {64}},\ \bibinfo {pages} {139} (\bibinfo {year}
  {2015})}\BibitemShut {NoStop}%
\bibitem [{\citenamefont {Wang}\ \emph {et~al.}(2020)\citenamefont {Wang},
  \citenamefont {Hu}, \citenamefont {Eggert}, \citenamefont {Fleischhauer},
  \citenamefont {Pelster},\ and\ \citenamefont {Zhang}}]{wang}%
  \BibitemOpen
  \bibfield  {author} {\bibinfo {author} {\bibfnamefont {T.}~\bibnamefont
  {Wang}}, \bibinfo {author} {\bibfnamefont {S.}~\bibnamefont {Hu}}, \bibinfo
  {author} {\bibfnamefont {S.}~\bibnamefont {Eggert}}, \bibinfo {author}
  {\bibfnamefont {M.}~\bibnamefont {Fleischhauer}}, \bibinfo {author}
  {\bibfnamefont {A.}~\bibnamefont {Pelster}},\ and\ \bibinfo {author}
  {\bibfnamefont {X.-F.}\ \bibnamefont {Zhang}},\ }\bibfield  {title} {\bibinfo
  {title} {Floquet-induced superfluidity with periodically modulated
  interactions of two-species hardcore bosons in a one-dimensional optical
  lattice},\ }\href {https://doi.org/10.1103/PhysRevResearch.2.013275}
  {\bibfield  {journal} {\bibinfo  {journal} {Phys. Rev. Res.}\ }\textbf
  {\bibinfo {volume} {2}},\ \bibinfo {pages} {013275} (\bibinfo {year}
  {2020})}\BibitemShut {NoStop}%
\bibitem [{\citenamefont {Goldman}\ and\ \citenamefont
  {Dalibard}(2014)}]{PhysRevX.4.031027}%
  \BibitemOpen
  \bibfield  {author} {\bibinfo {author} {\bibfnamefont {N.}~\bibnamefont
  {Goldman}}\ and\ \bibinfo {author} {\bibfnamefont {J.}~\bibnamefont
  {Dalibard}},\ }\bibfield  {title} {\bibinfo {title} {Periodically driven
  quantum systems: Effective {H}amiltonians and engineered gauge fields},\
  }\href {https://doi.org/10.1103/PhysRevX.4.031027} {\bibfield  {journal}
  {\bibinfo  {journal} {Phys. Rev. X}\ }\textbf {\bibinfo {volume} {4}},\
  \bibinfo {pages} {031027} (\bibinfo {year} {2014})}\BibitemShut {NoStop}%
\end{thebibliography}%

\appendix

\section{Appendix A: Derivation of Eq.~\eqref{eq:t_n recursion}}
For a general Floquet component $|\phi_n(j)\rangle\!=\! \phi_{j,n} |j\rangle$ Eq.~\eqref{eq:H_floquet} yields a set of coupled equations with
$\epsilon_n\!=\!\epsilon_0+n\omega$
\begin{align} \label{a1}
-J(\phi_{j+1,n}+\phi_{j-1,n})&=\epsilon_n \phi_{j,n}, \ \ |j|>1\\
\label{eq:Feq_phi2}
-J\phi_{\pm 2,n}-J'\phi_{0,n}&=\epsilon_n\phi_{\pm 1,n}\\
\nonumber
-J'(\phi_{1,n}+\phi_{-1,n})-\frac{\mu}{2}(\phi_{0,n-1}& +\phi_{0,n+1})\\\label{eq:Feq_phi3}
&=(\epsilon_n+\mu_0)\phi_{0,n}.
\end{align}
As expressed in  Eq.~\eqref{eq:Fansatz} we make the ansatz
\begin{align*}
\phi_{j,n}=\left\lbrace \begin{array}{cc}
t_{n}\mathrm{e}^{\mathrm{i}k_nj},&\qquad j> 0 \\
\xi_n ,&\qquad j=0 \\
\delta_{n,0}\mathrm{e}^{\mathrm{i}k_nj} + r_{n}\mathrm{e}^{-\mathrm{i}k_nj},& \qquad j<0
\end{array}\right..
\end{align*}
which gives $-2J\cos(k_n)\!=\!\epsilon_0\!+\!n\omega$ from Eq.~(\ref{a1}).   Using this dispersion and inserting the ansatz into Eq.~\eqref{eq:Feq_phi2} yieldswhich gives $-2J\cos(k_n)\!=\!\epsilon_0\!+\!n\omega$ from Eq.~(\ref{a1}).   Using this dispersion and inserting the ansatz into Eq.~\eqref{eq:Feq_phi2} yields

\begin{align}
\delta_{n,0}+r_{n}=t_{n}=\frac{J'}{J}\xi_n .
\label{eq:tn_rn}
\end{align}
Finally using relations \eqref{eq:Feq_phi3} and \eqref{eq:tn_rn} we are left with a set of algebraic equations as given by Eq. \eqref{eq:t_n recursion}.

\section{Appendix B: Derivation of Eq.~\eqref{impart}}
We want to calculate $\mathrm{Im}(c_0^++ c_0^-)$. For simplicity we define the small parameter $\lambda=\frac{J'^2}{J^2}$. 
Starting to evaluate the recursion Eq.~(\ref{eq:CF_rec})  for $c_n^+$  from a large cutoff value $M$, the expression
remains real if $\gamma_{n+1}\! \in\! \mathbb{R} $ 
until we reach small enough values of $n$.  This defines a second cutoff $N$ with $\rm{Im} c_{n\geq N}^+\!=\! 0$,  so only the Floquet states $n\!+\!1\!\leq\! N$ with $v_{n+1}\! \in\! \mathbb{R}$ are within the band, i.e.~$\epsilon_{n+1} \! \equiv\!  \epsilon_0\!+\! ({n\!+\!1})\omega\!\leq\!2J$, which creates a small imaginary part 
$\mathrm{Im}c_n^\pm = \mathcal{O}(\lambda)$. In particular, Eq.~\eqref{eq:CF_rec} yields for $v_{n+1} \in \mathbb{R}$
\begin{align}
    \mathrm{Im}c_n^+ &  =  \frac{\mu}{4}\frac{\lambda v_{n+1} + \mathrm{Im}c_{n+1}^+}{(\lambda v_{n+1}+\mathrm{Im}c_{n+1}^+)^2+(\mathrm{Re}c_{n+1} + \tilde{\epsilon}_{n+1}-\mu_0)^2} \nonumber \\
&\approx \frac{\mu}{4}\frac{\lambda v_{n+1} + \mathrm{Im}c_{n+1}^+}{(\mathrm{Re}c_{n+1} + \tilde{\epsilon_0} +(n+1)\tilde{\omega}-\mu_0)^2} \nonumber \\
&\approx  \frac{4}{\mu^2}(-\mathrm{Re}c_n^+)^2(\lambda v_{n+1}+\mathrm{Im}c_{n+1}^+) \label{im}
\end{align}
where we have omitted higher order terms in $\lambda$ and used 
\begin{align}
     \mathrm{Re}c_n^\pm \approx -\frac{\mu^2}{4}\frac{1}{\mathrm{Re}c_{n+1}^\pm+\tilde{\epsilon_0} \pm(n+1)\tilde{\omega}-\mu_0 }
\end{align}
from Eq.~\eqref{eq:CF_rec}. As argued above, $\mathrm{Im}c_{N}^+=0$ is the starting point in Eq.~(\ref{im}), so the expression can be iterated in a straight-forward way to evaluate
\begin{align}
\nonumber
\mathrm{Im}c_0^+ &= \lambda \sum_{n=1}^N v_n\left(\prod_{j=0}^{n-1}\frac{4}{\mu^2}(\mathrm{Re}c_j^+)^2\right) + \mathcal{O}(\lambda^2)\\
&=\lambda \sum_{n=1}^N v_n\frac{\mathcal{J}_{\nu_0+n}(\nicefrac{\mu}{\tilde{\omega}})^2}{\mathcal{J}_{\nu_0}(\nicefrac{\mu}{\tilde{\omega}})^2} + \mathcal{O}(\lambda^2)
\label{eq:c+}
\end{align}
where Eqs.~(\ref{bessel}) and \eqref{a_n}  were used on the last line.
Analogously, we find
\begin{align}
\nonumber
\mathrm{Im}c_0^- &=\lambda \sum_{n=1}^{N'} v_{-n}\left(\prod_{j=0}^{n-1}\frac{4}{\mu^2}(\mathrm{Re}c_{j}^-)^2\right) + \mathcal{O}(\lambda^2)\\
&=\lambda \sum_{n=1}^{N'} v_{-n}\frac{\mathcal{J}_{-\nu_0+n}(\nicefrac{\mu}{\tilde{\omega}})^2}{\mathcal{J}_{-\nu_0}(\nicefrac{\mu}{\tilde{\omega}})^2} + \mathcal{O}(\lambda^2)
\label{eq:c-}
\end{align}
where the cutoff $N'$ is defined so that for $n\!+\!1\!\leq\!N'$ all $v_{-(n+1)}\! \in\! \mathbb{R}$, i.e.~when Floquet components enter the band from below   
$  \epsilon_0\!-\! ({n\!+\!1})\omega\!\geq\!-2J$.   

Finally, note that if $|\epsilon_0\pm\omega|\geq 2J$ it follows $\mathrm{Im}c_0^\pm = 0$ independent of any approximation. 
Therefore, if no Floquet components are located within the energy band, we can 
find perfect transmission by setting the real part to zero, which can be always achieved as a function of $\omega$, $\mu$, and $J'$, translating into a resonance
condition $\omega_{\rm res} (\mu, J')$.
\end{document}